\documentclass[reprint,aps,superscriptaddress]{revtex4-2}
\usepackage{amsmath}
\usepackage{amssymb}
\usepackage{graphicx}
\usepackage[caption=false]{subfig}
\usepackage{float}
\usepackage{bm}
\usepackage{indentfirst}
\usepackage[colorlinks=true,linkcolor=blue,citecolor=blue,urlcolor=blue]{hyperref}
\usepackage[usenames,dvipsnames]{xcolor}
\usepackage{siunitx}
\usepackage{color}

\begin{document}
\title{In-plane magnetic field induced density wave states \\ near quantum spin Hall phase transitions}
\author{Yongxin Zeng}   
\affiliation{Department of Physics, The University of Texas at Austin, Austin, TX 78712}
\author{Fei Xue}
\affiliation{Department of Physics, University of Alabama at Birmingham, Birmingham, AL 35294}
\author{Allan H. MacDonald}
\affiliation{Department of Physics, The University of Texas at Austin, Austin, TX 78712}
\date{\today}
\begin{abstract}
We study the influence of an in-plane magnetic field and Coulomb interactions on the physics of quantum spin Hall insulators, like those  
in InAs/GaSb and HgTe/CdTe quantum wells.  Using a Hartree-Fock mean-field theory approximation, we calculate phase diagrams as functions of the band gap, band hybridization, and magnetic field strength. We show that when the band hybridization is weak, the system is unstable against the 
formation of density wave states. As the strength of the in-plane magnetic field increases, the density-wave
region of the phase diagram expands and distinct density-wave states appear. 
We discuss possible experimental implications of our results.
\end{abstract}
\maketitle

\section{Introduction}

The quantum spin Hall (QSH) insulator is a topologically nontrivial state of matter characterized by gapless helical edge states protected by time-reversal symmetry \cite{KaneMele_QSH,KaneMele_Z2}. It was first realized in HgTe/CdTe quantum wells \cite{bernevig2006quantum,konig2007quantum}, and later also in other systems like InAs/GaSb quantum wells \cite{liu2008quantum,knez2011evidence,du2015robust}. Theoretically the physics of QSH insulators is captured by the 
Bernevig-Hughes-Zhang (BHZ) model \cite{bernevig2006quantum}, which is a single-particle theory that 
ignores interactions and works well in the limit of strong band hybridization.
Interactions become important when the BHZ model band hybridization parameter $A$ is small, as can be appreciated by 
considering the limit $A \to 0$, where coherence between conduction and valence bands, or exciton condensation,
occurs spontaneously when the band gap is smaller than the exciton binding energy \cite{Keldysh1965,jerome1967excitonic,Lozovik1976,Comte1982,zhu1995exciton,lozovik1996phase,high2012spontaneous,wu2015theory}. 
Recent experiments \cite{du2017evidence,wu2019electrically,wu2019resistive} have shown excitonic behavior in InAs/GaSb quantum wells. 
The interplay between interactions and topology can lead to interesting new phases near the QSH phase transition \cite{budich2014time,pikulin2014interplay,hu2017topological,xue2018time}, which have so far been only lightly explored.

In this paper we study how an in-plane magnetic field modifies the phase diagram studied in Ref.~\onlinecite{xue2018time},
which contains time-reversal symmetry-breaking electron nematic phases.  Due to the spatial separation between electron and hole layers, an in-plane magnetic field shifts the conduction and valence bands in opposite directions in momentum space. 
Intuitively this opposite shift effectively increases the band gap and reduces the hybridization between the electron and hole bands.
When interactions are neglected, an in-plane magnetic field drives the system into a semi-metallic state.
Using a Hartree-Fock mean-field theory, we show that the nematic states instead break translational symmetry and become density-wave states.  At stronger tunneling an in-plane magnetic field can
drive the system through a variety of different phases, including quantum anomalous Hall states with and without
density-wave order.  

This paper is organized as follows: In Section \ref{sec:mft} we formulate the mean-field theory we use to describe interaction effects, 
and explain how we allow the possibility of translational symmetry breaking. 
In Section \ref{sec:phase} we summarize our results by presenting phase diagrams 
that depend on three parameters: band gap, hybridization, and the strength of in-plane magnetic field. 
Finally in Section \ref{sec:discussion} we discuss the relationship between our work and potential 
future experiments, and its relationship to excitonic density-wave states that have been identified, 
often controversially, in bulk three-dimensional crystals.  

\section{Mean-field theory} \label{sec:mft}

We use a four-band BHZ model \cite{bernevig2006quantum,liu2008quantum} to describe the InAs/GaSb quantum wells. 
The field operators are four-component spinors $\psi_{\bm k}=(a_{c\uparrow\bm k},a_{v\uparrow\bm k},a_{c\downarrow\bm k},a_{v\downarrow\bm k})^T$, where $c$ and $v$ denote the conduction and valence bands, and $\uparrow$ and $\downarrow$ denote two opposite spins.
The single-particle physics of the system under an in-plane magnetic field is described by the modified BHZ 
Hamiltonian \cite{Hu2016,hu2017topological}
\begin{equation} \label{eq:H_BHZ}
H_{\rm{BHZ}} = \sum_{\bm k} \psi_{\bm k}^{\dagger} \left(\begin{array}{cc}
h_{\uparrow}(\bm k) & 0 \\
0 & h_{\downarrow}(\bm k)
\end{array}\right) \psi_{\bm k},
\end{equation}
where the two $2\times 2$ matrices $h_{\uparrow}$ and $h_{\downarrow}$ can be explicitly expressed as
\begin{equation} \label{eq:h}
\begin{split}
h_{\uparrow}(\bm k) &= \left(\begin{array}{cc}
\frac{\hbar^2}{2m_e}(\bm k-\frac{\bm Q}{2})^2+\frac{E_g}{2} & A(k_x+ik_y) \\
A(k_x-ik_y) & -\frac{\hbar^2}{2m_h}(\bm k+\frac{\bm Q}{2})^2-\frac{E_g}{2}
\end{array}\right), \\
h_{\downarrow}(\bm k) &= \left(\begin{array}{cc}
\frac{\hbar^2}{2m_e}(\bm k-\frac{\bm Q}{2})^2+\frac{E_g}{2} & -A(k_x-ik_y) \\
-A(k_x+ik_y) & -\frac{\hbar^2}{2m_h}(\bm k+\frac{\bm Q}{2})^2-\frac{E_g}{2}
\end{array}\right).
\end{split}
\end{equation}
$m_e$ and $m_h$ are the effective masses of electrons and holes, $E_g$ is the band gap, $A$ is the strength of hybridization between the conduction and valence bands, and $\bm Q$ is the momentum shift due to the in-plane magnetic field. Without the magnetic field, $\bm Q=0$ and $h_{\uparrow},h_{\downarrow}$ are time-reversal partners:
\begin{equation}
h_{\uparrow}(\bm k) = h_{\downarrow}^*(-\bm k).
\end{equation}
When an in-plane magnetic field $\bm B=B\hat y$ is applied to electron and hole layers separated by an interlayer distance $d$, the conduction and valence bands are shifted in momentum by $\mp \bm Q=\mp (eBd/\hbar) \hat x$ by Peierls substitution. 
The in-plane magnetic field breaks the time-reversal symmetry of the system and 
induces orbital moments.  The electrons and holes interact via the Coulomb interaction
\begin{equation}
H_I = \frac{1}{2S}\sum_{bb'ss'}\sum_{\bm{kk'q}} V_{bb'}(q) a_{bs\bm k+\bm q}^{\dagger} a_{b's'\bm k'-\bm q}^{\dagger} a_{b's'\bm k'} a_{bs\bm k},
\end{equation}
where $S$ is the area of the two-dimensional system, $b,b'$ and $s,s'$ are band and spin indices respectively, $V_{cc}(q)=V_{vv}(q)=V(q)=2\pi e^2/\epsilon q$ is the intralayer Coulomb interaction, $V_{cv}(q)=V_{vc}(q)=U(q)=V(q)\exp(-qd)$ is the interlayer Coulomb interaction at interlayer distance $d$, and $\epsilon$ is the dielectric constant of the surrounding three-dimensional material.

Anticipating the possibility of translational symmetry breaking \cite{hu2017topological} along the 
direction of $\bm{Q}$, we divide momentum space into slabs separated by $\bm{Q}$. 
Then apart from the band and spin indices $(b,s)$, the basis states are labeled by an integer $n$ and a quasi-momentum $\bm k$ ($|\bm{k}|<|\bm{Q}|/2$) that lies within the first quasi-one-dimensional Brillouin zone.
Together $n$ and $\bm{k}$ refer to the plane-wave state with momentum $n\bm Q+\bm k$.

We use a Hartree-Fock mean-field theory to describe the Coulomb interaction. The Hartree term is
\begin{equation}
\begin{split}
\Sigma_H = \frac{1}{S}\sum_{bb'\atop ss'}\sum_{nn'n''\atop \bm k\bm k'} &V_{bb'}((n'-n)\bm Q) \\
\times&\rho_{b's'\,n''}^{b's'\,n''+n'-n}(\bm k') \; a_{bsn'\bm k}^{\dagger} a_{bsn\bm k},
\end{split}
\end{equation}
where the density matrix $\rho$ is defined relative to the density matrix with valence bands filled and conduction bands empty:
\begin{equation} \label{eq:rho}
\rho_{b's'n'}^{bsn}(\bm k) = \langle a_{b's'n'\bm k}^{\dagger} a_{bsn\bm k} \rangle - \delta_{bb'}\delta_{bv}\delta_{ss'}\delta_{nn'}.
\end{equation}
For $n'=n$, the Hartree term accounts for the electrostatic potential energy difference $4\pi e^2 n_x d/\epsilon$ between the electron and hole layers, where
\begin{equation}
n_x = \frac{1}{S} \sum_{sn\bm k} \rho_{csn}^{csn}(\bm k) = -\frac{1}{S} \sum_{sn\bm k} \rho_{vsn}^{vsn}(\bm k)
\end{equation}
is the exciton density. The Fock term
\begin{equation}
\begin{split}
\Sigma_F = -\frac{1}{S}\sum_{bb'\atop ss'}\sum_{nn'n''\atop \bm k\bm k'} &V_{bb'}((n''-n)\bm Q+\bm k'-\bm k) \\
\times&\rho_{bs\,n''}^{b's'\,n''+n'-n}(\bm k') a_{b's'n'\bm k}^{\dagger} a_{bsn\bm k}.
\end{split}
\end{equation}
accounts for the exchange interaction. Together, the system is described by the mean-field Hamiltonian
\begin{equation} \label{eq:H_MF}
H_{\rm{MF}} = H_{\rm{BHZ}} + \Sigma_H + \Sigma_F.
\end{equation}

Below we express lengths and energies in terms of characteristic scales $a_B^* = \epsilon\hbar^2/me^2$ and $Ry^* = e^2/2\epsilon a_B^*$, where $m=m_e m_h/(m_e+m_h)$ is the reduced effective mass. This model approximates InAs/GaSb quantum wells if we choose $m_e=0.023m_e,m_h=0.4m_0$, and $\epsilon=15$ \cite{levinshtein1996handbook}, which implies that $a_B^*=\SI{36.5}{nm}$ and $Ry^*=\SI{1.3}{meV}$. We assume the interlayer distance $d=0.3~a_B^*\approx\SI{10}{nm}$. For an in-plane magnetic field of strength $B=\SI{1}{T}$, the momentum shift $Q=0.606~(a_B^*)^{-1}$. For simplicity in Eq.~\eqref{eq:h} we neglect the particle-hole asymmetry which does not affect the ground state physics, and perform numerical calculations with $m_e=m_h=2m$.

\section{Phase diagrams} \label{sec:phase}

\begin{figure*}
\centering
\subfloat{\includegraphics[width=0.48\linewidth]{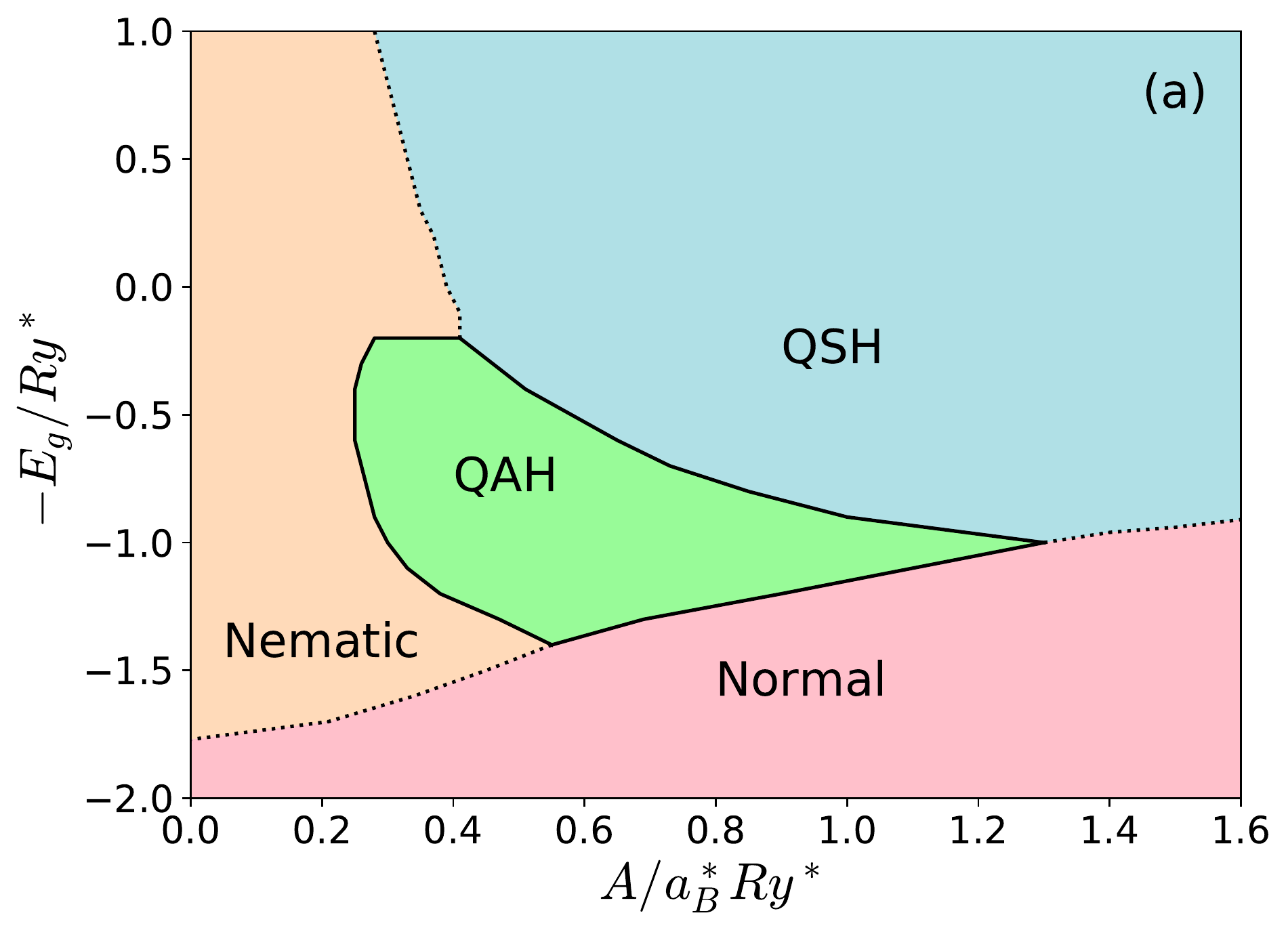}}
\subfloat{\includegraphics[width=0.48\linewidth]{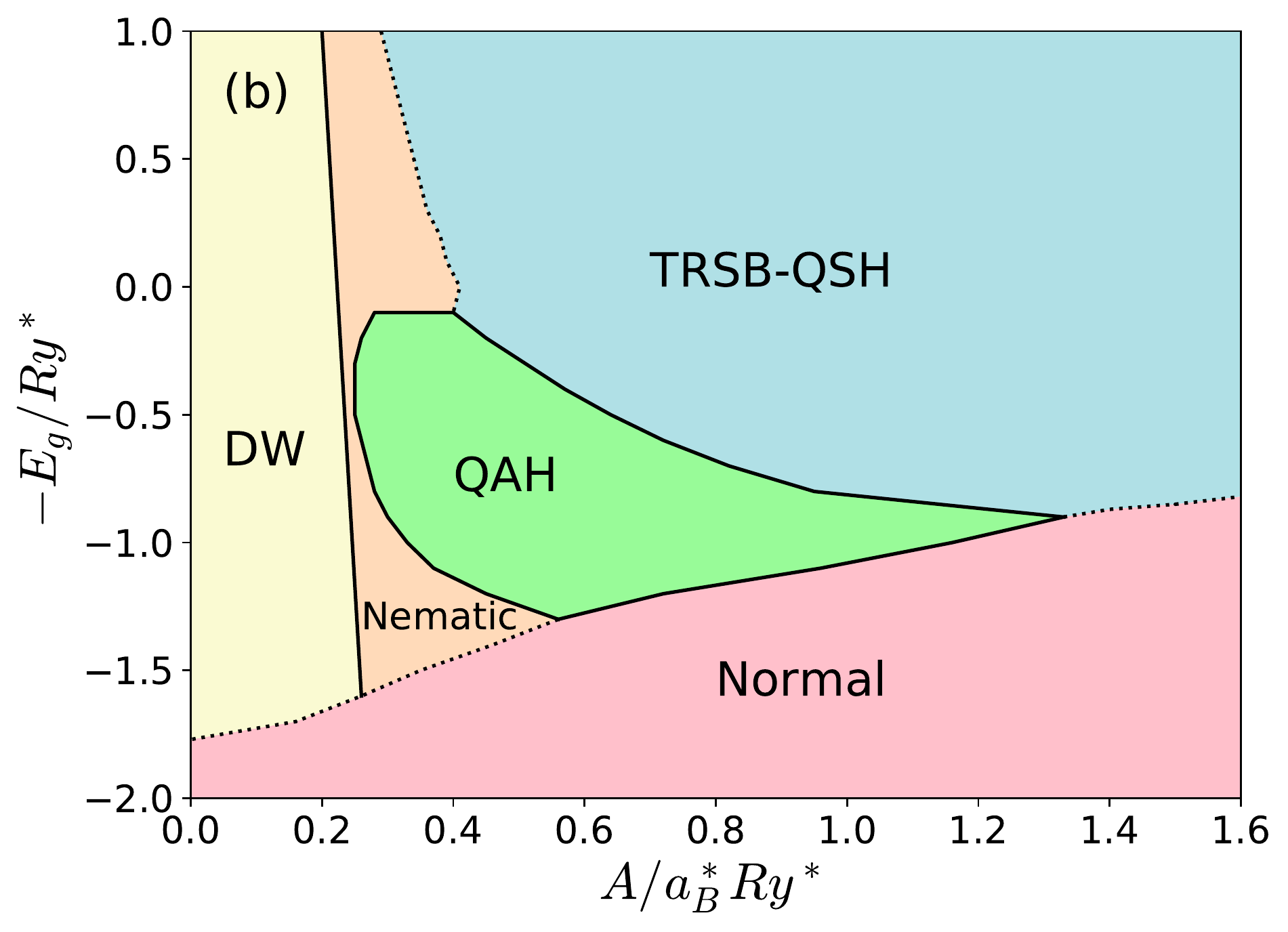}}\\
\subfloat{\includegraphics[width=0.48\linewidth]{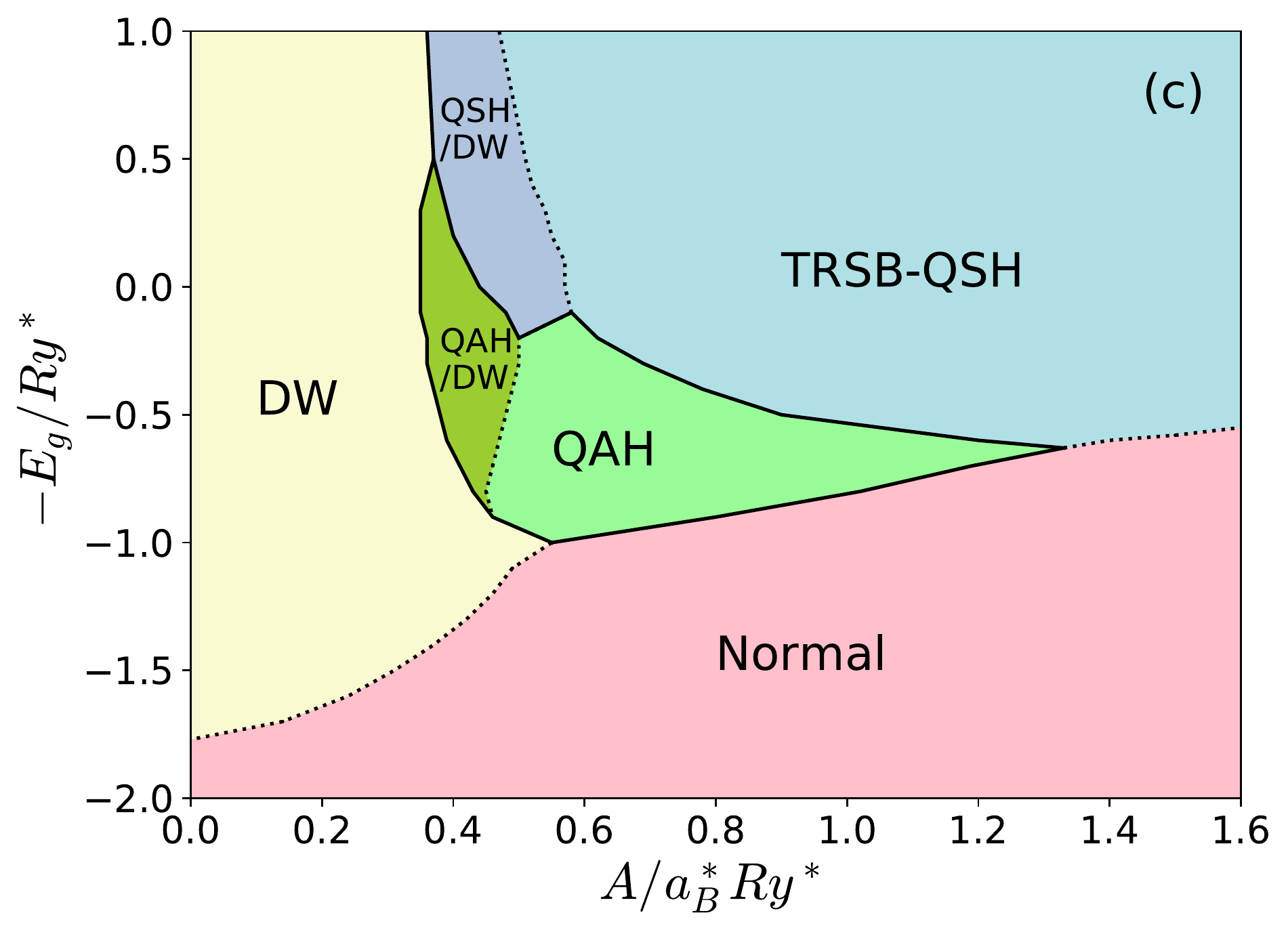}}
\subfloat{\includegraphics[width=0.48\linewidth]{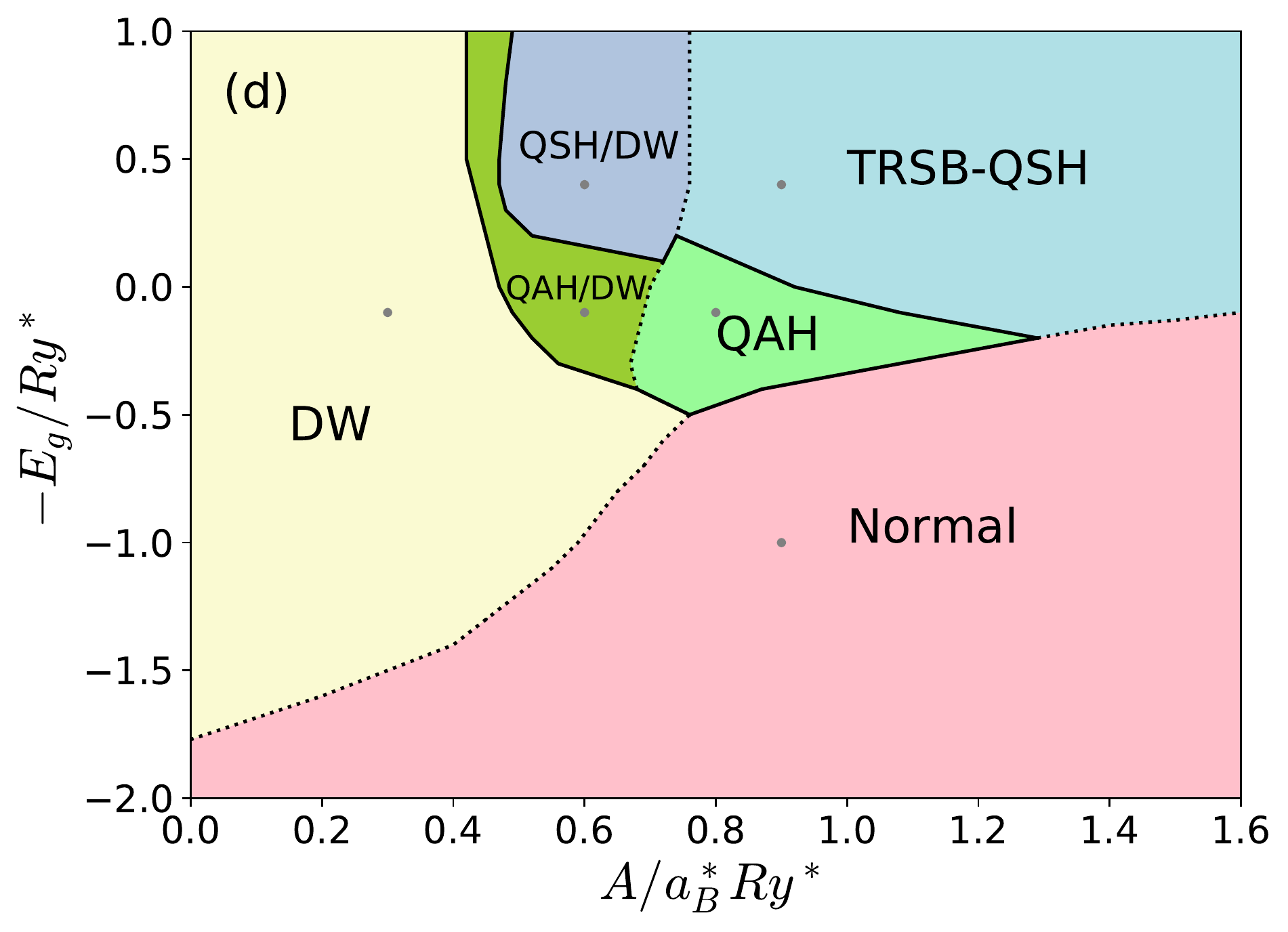}}
\caption{Phase diagrams in $(A,E_g)$ plane at several fixed magnetic fields: (a) $Q=0$; (b) $Qa_B^*=0.6$; (c) $Qa_B^*=1.2$; (d) $Qa_B^*=1.8$. Solid and dotted lines distinguish first-order and continuous phase transitions. The gray dots in (d) stand for the parameters used for later illustrative calculations in Figs.~\ref{fig:charge_density} and \ref{fig:wilson}. For adjacent InAs and GaSb layers with thickness $d_{\rm InAs} = d_{\rm GaSb} = \SI{10}{nm}$, the hybridization strength \cite{liu2013models} $A = \SI{0.37}{eV.\AA} = 0.78\, a_B^* Ry^*$.
The properties of the different phases identified here are described in the main text.} \label{fig:phase_fixQ}
\end{figure*}

Given $E_g$, $A$ and $Q$, the ground state of the system can be obtained by solving the Hartree-Fock equations self-consistently and finding the lowest-energy solution. The $Q=0$ case has been studied in detail in Ref.~\onlinecite{xue2018time}, and the phase diagram is reproduced in Fig.~\ref{fig:phase_fixQ}(a). At $A=0$ the 
number of particles in the conduction and valence bands are conserved separately. When the bare energy 
gap $E_g$ is reduced below the $1s$ exciton binding energy, the excitons condense and 
coherence is established spontaneously between the conduction and valence bands.
This order survives at finite $A$, where it breaks rotational symmetry by establishing 
coherence between $s$-conduction and $p$-valence electrons, and also breaks time-reversal symmetry by doing so in a 
spin-dependent manner.
In Fig.~\ref{fig:phase_fixQ} we refer to this state as the nematic insulator state. 
At large $A$, single-particle physics dominates and the system undergoes a topological phase transition between the QSH and normal insulators as $E_g$ varies. At moderate values of $A$, the transition between QSH and normal insulators
occurs via an intermediate quantum anomalous Hall state (QAH) state with 
broken time-reversal symmetry and a nonzero Chern number \footnote{In this summary, 
we have ignored some minor phases that occur in a small region of the phase diagram near the $A=0$ line.}.

When an in-plane magnetic field is applied, the QSH state is no longer protected by time-reversal symmetry. 
In the simplified BHZ Hamiltonian \eqref{eq:H_BHZ} we use here, the two spins are decoupled and a spin Chern number can be defined to distinguish QSH and normal insulators \cite{sheng2006quantum}. In the more general case where spin is not a good quantum number, it has been shown that \cite{prodan2009robustness,yang2011time,vanderbilt2018berry} the spin Chern number can remain well-defined as a robust topological invariant. For this reason the QSH-normal insulator transition still exists.  In order to distinguish these two cases 
we refer to the finite-$B$ QSH state as a time-reversal symmetry-breaking (TRSB) QSH state \cite{yang2011time}.

At small hybridization $A$ the momentum-shifted conduction and valence bands tend to establish coherence by breaking translational symmetry to achieve better Fermi surface nesting, forming density wave (DW) states with wavevector $\bm Q$. 
We find that at small but finite $A$ the energetically preferred state is one in which
pairing is between opposite spins, so the order parameter is $\rho_{v\bar s 0}^{cs1}(0)$ where $\bar{s}$ denotes the spin 
opposite to $s$.
At $A=0$ the density matrix element $\rho_{b'\bar s 0}^{bs1}$ is nonzero only for $b=c,b'=v$.
When the band-hybridization parameter $A$ is non-zero, on the the other hand, 
it is nonzero for any $b$ and $b'$, although the exciton condensate order parameter ($b=c,b'=v$) is always much larger than the other three ($b=b'=v$, $b=b'=c$, and $b=v,b'=c$) density-matrix elements.
When the magnetic field is weak, the DW state exists only near $A=0$ and undergoes a first-order phase transition to the nematic insulator state, which does not have finite-$Q$ pairing, as $A$ increases (see Fig.~\ref{fig:phase_fixQ}(b)). 
The nematic insulator phase is characterized by the order parameter $\rho_{v\bar s 0}^{cs0}(0)$; the coherence that
is established does not accomodate the momentum-space shifts of the conduction and valence bands.
We retain the term nematic insulator used at $Q=0$ even though rotational symmetry has already been explicitly 
broken by the in-plane magnetic field.

The boundary between the magnetic-field-stabilized DW state and the nematic insulator 
state moves rapidly to the right as the magnetic field strength increases, eventually 
squeezing the nematic insulator state out of the phase diagram as shown in Fig.~\ref{fig:phase_fixQ}(c). 
Near the DW phase boundary neighboring the TRSB-QSH and QAH phase regions, two new phases appear \footnote{We have ignored some minor phases that appear near the phase boundaries. We find that the QSH/DW and QAH/DW phases are more stable as the magnetic field strength increases.} that also break translational symmetry along the $\hat x$-direction. These two states are connected to the TRSB-QSH and QAH states via continuous phase transitions, and we label them as QSH/DW and QAH/DW states respectively, for reasons that will become clear later. Like the DW state, both the QSH/DW and QAH/DW states have order parameters of the form $\rho_{b'\bar s 0}^{bs1}(0)$. Unlike the DW state, however, the largest pairing terms in the QSH/DW state are between conduction and conduction, and valence and valence bands, yielding order parameters $\rho_{v\bar{s}0}^{vs1}(0)(=\rho_{c\bar{s}\bar{1}}^{cs0}(0))$. The QAH/DW state has different up-to-down and down-to-up spin pairings: $\rho_{b'\uparrow 0}^{b\downarrow 1} \ne \rho_{b'\downarrow 0}^{b\uparrow 1}$, with one, say $\rho_{b'\uparrow 0}^{b\downarrow 1}$, resembling the DW state with the largest element appearing at $b=c,b'=v$, and the other ($\rho_{b'\downarrow 0}^{b\uparrow 1}$)
resembling the QSH/DW state with the largest element at $b=b'=v$.
The transitions between the DW, QAH/DW and QSH/DW states are all first-order transitions.

As $Q$ continues to increase, the three density-wave regions keep expanding and the QAH region shrinks as shown in Fig.~\ref{fig:phase_fixQ}(d).
While the transitions between the TRSB-QSH, QAH and normal insulator states stay largely unchanged, the phase boundaries slowly move towards larger $A$ and smaller $E_g$ as $Q$ increases, consistent with the intuition that the momentum shift between conduction and valence bands effectively increases the band gap and weakens the band hybridization.


\begin{figure}
\centering
\includegraphics[width=\linewidth]{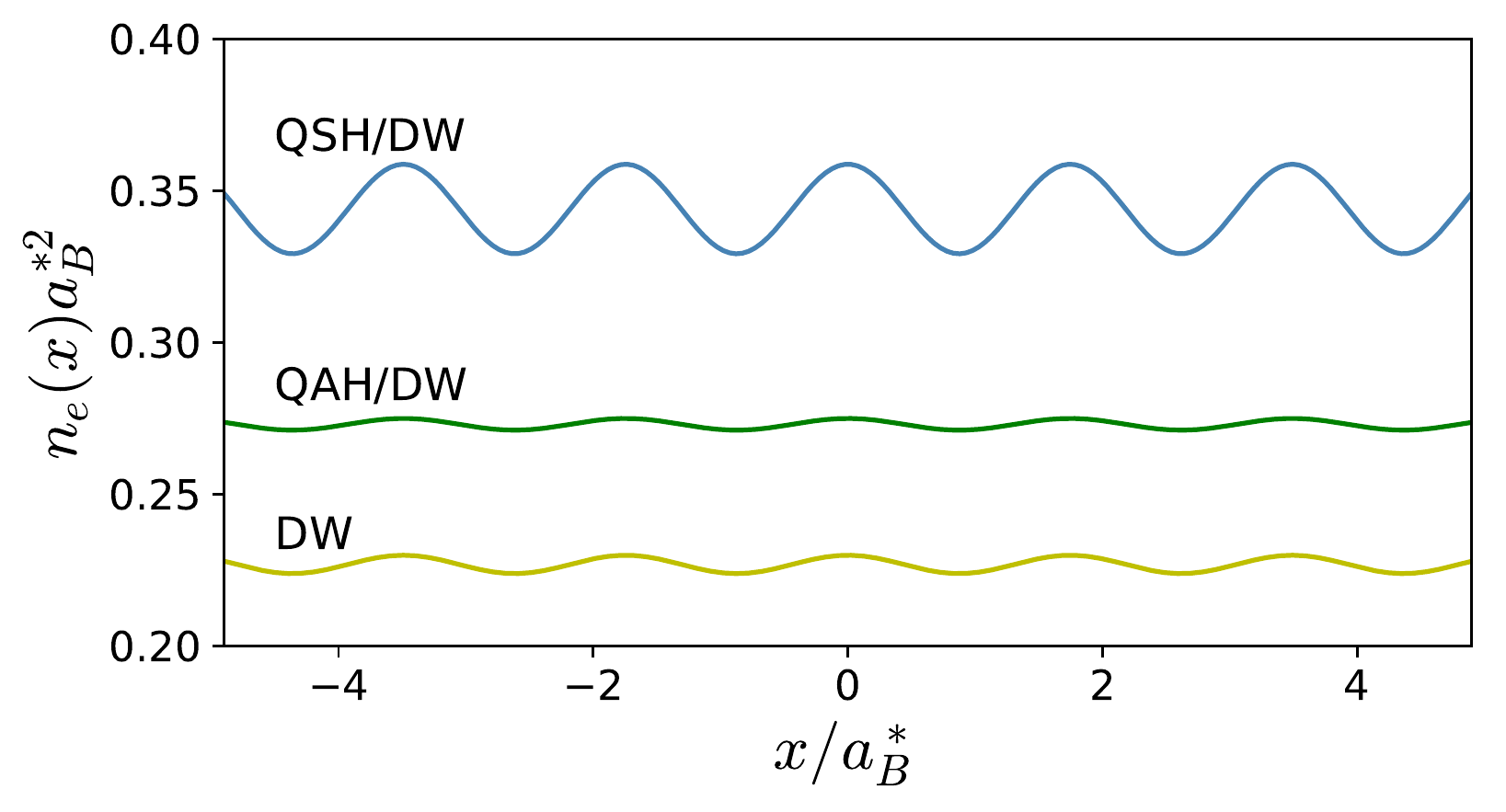}
\caption{Spatial distribution of the charge density (in units of $-e$) in the electron layer $n_e$ at $Qa_B^*=1.8$ and (i) $E_g=0.1\,Ry^*, A=0.3\,a_B^* Ry^*$ (DW, yellow); (ii) $E_g=0.1\,Ry^*, A=0.5\,a_B^* Ry^*$ (QAH/DW, green); (iii) $E_g=-0.4\,Ry^*, A=0.5\,a_B^* Ry^*$ (QSH/DW, blue).} \label{fig:charge_density}
\end{figure}

For $A=0$, the ground states are true exciton condensates and ordering does not lead to charge-density variations in 
either layer even if it occurs at $Q \ne 0$.  For $A \ne 0$, the situation changes. 
The charge densities in the electron and hole layers (in units of $-e$) are related to the density matrices $\rho$ by 
\begin{equation}
\begin{split}
n_e(\bm r) &= \frac 1S \sum_{snn'\bm k} \rho_{csn'}^{csn}(\bm k) e^{-i(n'-n)\bm Q\cdot\bm r}, \\
n_h(\bm r) &= \frac 1S \sum_{snn'\bm k} \rho_{vsn'}^{vsn}(\bm k) e^{-i(n'-n)\bm Q\cdot\bm r}.
\end{split}
\label{eq:density}
\end{equation}
At $A=0$, the only nonzero $n' \ne n$ density-matrix elements in Eq.~\eqref{eq:density} are $\rho_{c\bar s\, n+1}^{vs\, n}$, 
so the charge density is uniform in each layer. When $A\ne 0$, $\rho^{bs\, n}_{bs\, n\pm 1}$ still vanishes because only opposite spins in neighboring Brillouin zones are coupled. However, $\rho^{bs\, n}_{bs\, n\pm 2}$ can be nonzero when $A\ne 0$, so the charge density in each layer oscillates periodically in space, with periodicity $\pi/Q$. Fig.~\ref{fig:charge_density} shows the charge density distribution in the electron layer of the three density-wave states. We see that the QSH/DW state has larger spatial charge density fluctuations than the other density-wave states, as expected because of its conduction-to-conduction and valence-to-valence band couplings. In our calculation with $m_e=m_h$, the charge density in the hole layer is exactly the opposite of that in the electron layer due to the particle-hole symmetry present for this parameter choice. 
In the general $m_e\ne m_h$ case there is partial cancellation between the charge densities in the two layers, resulting in a weak total charge density oscillation in space. 

\begin{figure}
\centering
\includegraphics[width=\linewidth]{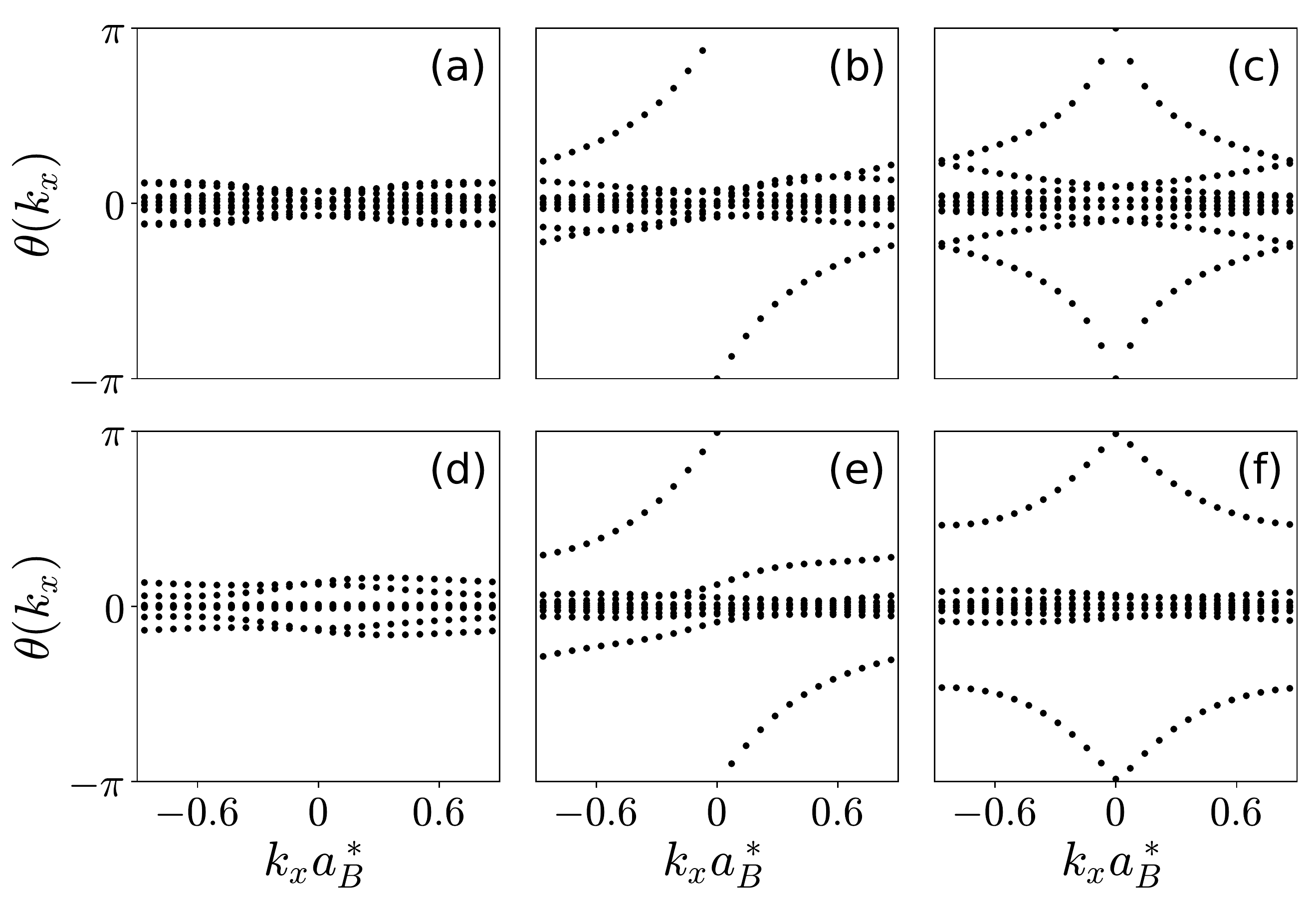}
\caption{Wilson loop calculations at (a) $E_g=1\,Ry^*, A=0.9\,a_B^*Ry^*$ (normal insulator); (b) $E_g=0.1\,Ry^*, A=0.8\,a_B^*Ry^*$ (QAH); (c) $E_g=-0.4\,Ry^*, A=0.9\,a_B^*Ry^*$ (TRSB-QSH); (d) $E_g=0.1\,Ry^*, A=0.3\,a_B^*Ry^*$ (DW); (e) $E_g=0.1\,Ry^*, A=0.6\,a_B^*Ry^*$ (QAH/DW); (f) $E_g=-0.4\,Ry^*, A=0.6\,a_B^*Ry^*$ (QSH/DW).} \label{fig:wilson}
\end{figure}

\begin{figure*}
\centering
\includegraphics[height=0.36\linewidth]{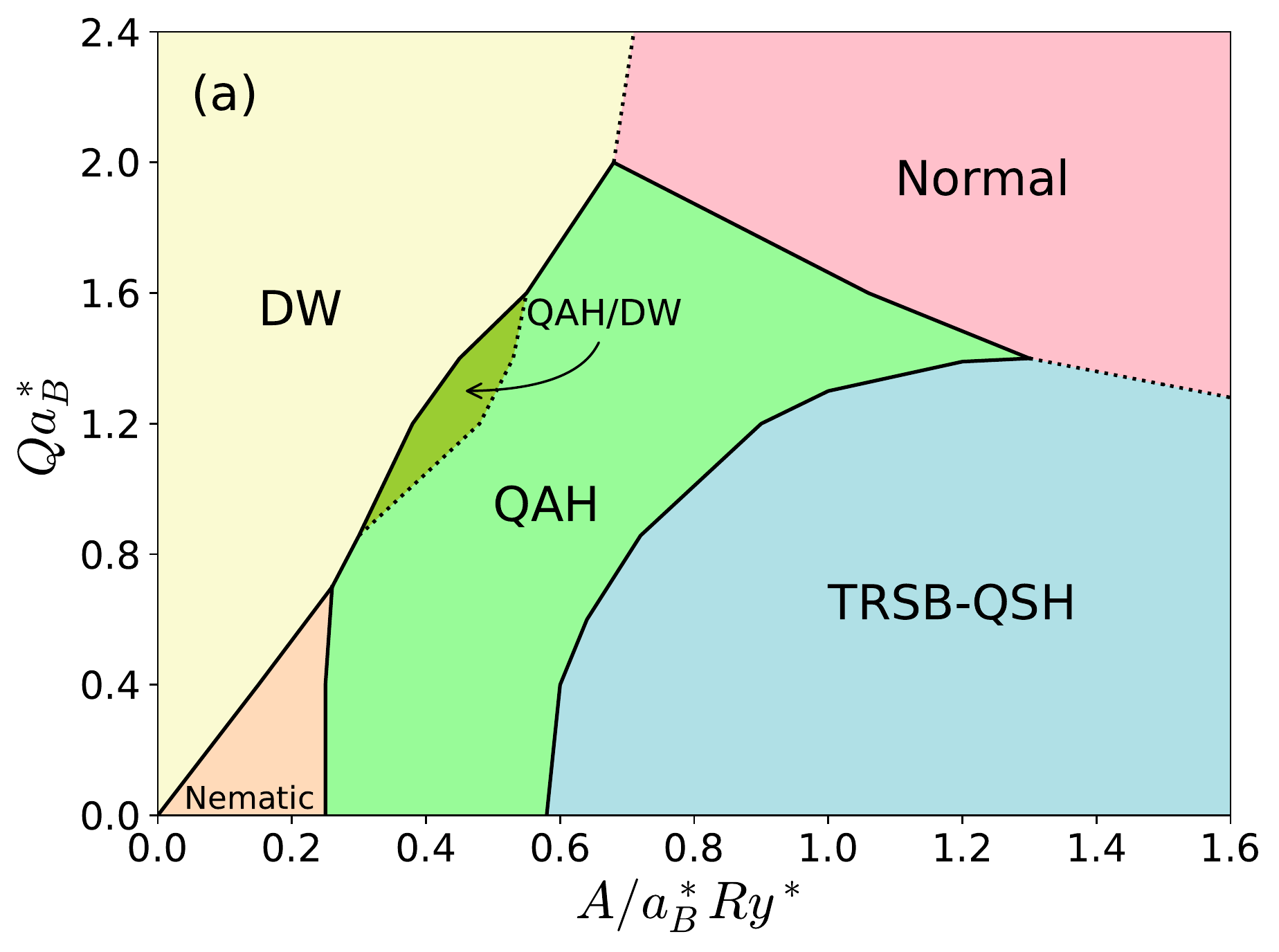}
\includegraphics[height=0.36\linewidth]{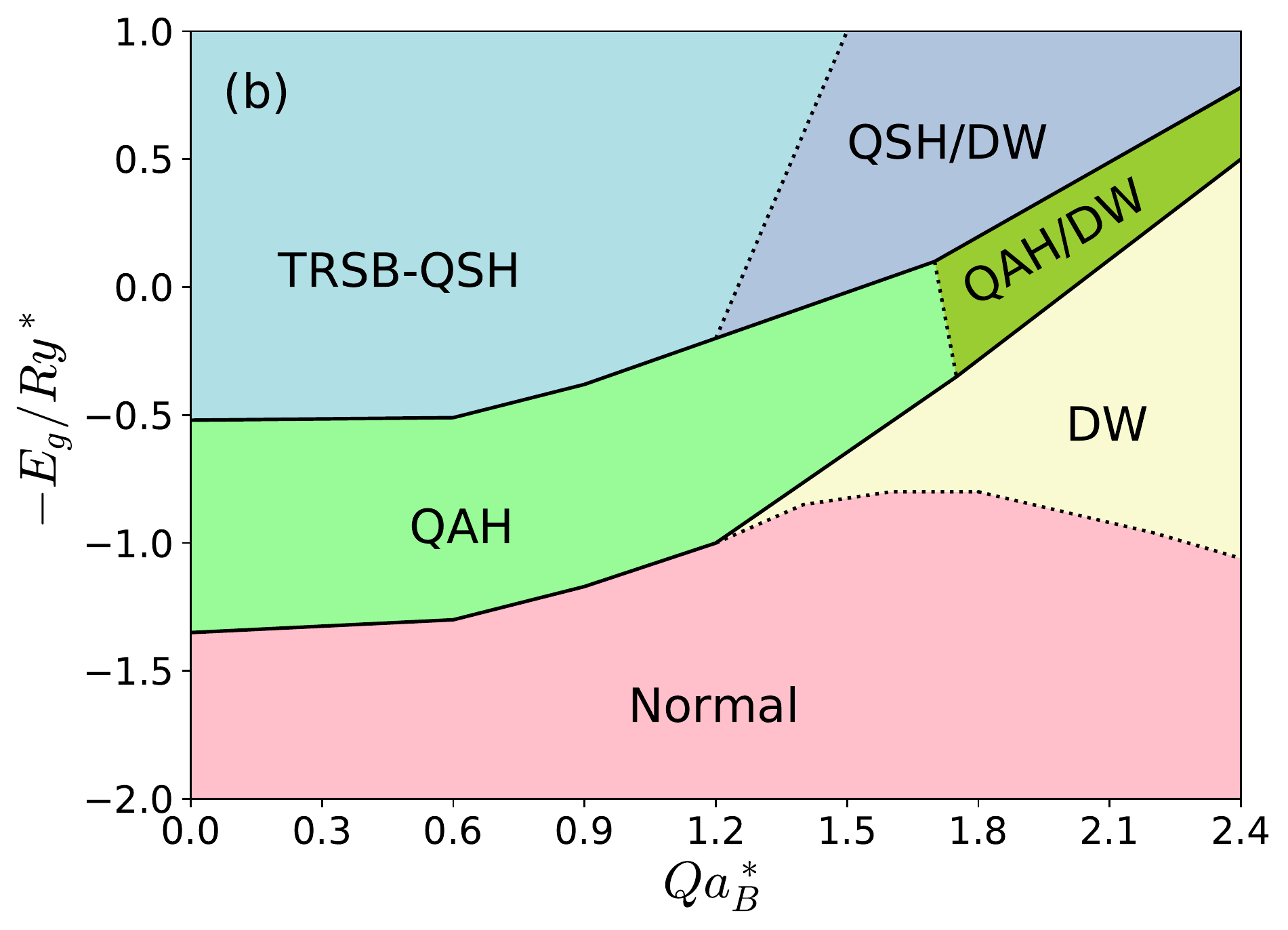}
\caption{Phase diagram (a) in $(A,Q)$ plane at fixed $E_g=0.5\,Ry^*$; (b) in $(Q,E_g)$ plane at fixed $A=0.6\,a_B^* Ry^*$.} \label{fig:phase_varyQ}
\end{figure*}

The topological properties of these phases can be studied by performing Wilson loop calculations \cite{yu2011equivalent,weng2015quantum}.  Our results are shown in Fig.~\ref{fig:wilson}. 
For each $k_x$, we calculate the product $D(k_x)$ of the Berry connection matrices $F_{i,i+1}^{m,n} = \langle u_i^m | u_{i+1}^n \rangle$ along $k_y$, where $i$ labels steps along $k_y$ and $m,n$ label occupied states.  
We then calculate the phase angles $\theta(k_x)$ of the eigenvalues of the matrix $D(k_x)$. The topological properties of the system can be read from the winding behavior of the phase angles. For example, the Chern number is equal to the net number of times (upwards minus downwards) the evolution curves of $\theta$ cross a constant reference line parallel to the $k_x$-axis. The results show that the phase angles of the normal insulator (Fig.~\ref{fig:wilson}(a)) and DW (Fig.~\ref{fig:wilson}(d)) states always stay near zero, so these two states are both topologically trivial. In the QAH state (Fig.~\ref{fig:wilson}(b)) one of the spins undergoes band inversion, and the phase angle $\theta$ jumps by $2\pi$ as $k_x$ sweeps from $-Q/2$ to $Q/2$. Interestingly, the Wilson loop of the QAH/DW state (Fig.~\ref{fig:wilson}(e)) shows very similar winding behavior as the QAH state. We conclude that both the QAH and QAH/DW states are topologically nontrivial, characterized by Chern number $|\mathcal C|=1$. For the TRSB-QSH state (Fig.~\ref{fig:wilson}(c)) the two decoupled spin bands are both inverted and the Wilson loops exhibit nontrivial but opposite winding along $k_x$, resulting in zero total Chern number. In the QSH/DW state (Fig.~\ref{fig:wilson}(f)), however, the two spins are coupled via density-wave order, and the degeneracy at $k_x=\pm Q/2$ is lifted. Despite certain similarity to that of the QSH state, the Wilson loop of the QSH/DW state suggests that the system is topologically trivial \footnote{In principle we may apply the definition of spin Chern number in Ref.~\onlinecite{prodan2009robustness}. However, here we follow a more conservative convention and do not calculate the spin Chern number for the QSH/DW state, because the definition involves a rather artificial requirement on adiabatic path \cite{vanderbilt2018berry} and lacks physical significance.}. In fact, the QSH/DW state we find is very similar to the topological charge density wave state discovered in Ref.~\onlinecite{hu2017topological}, except that in Ref.~\onlinecite{hu2017topological} the density-wave order parameter is between the same spins, whereas we find the system has lower energy when the coupling is between opposite spins. It is the coupling between different spins that lifts the degeneracy at $k_x=\pm Q/2$ and gives rise to a trivial Wilson loop as shown in Fig.~\ref{fig:wilson}(f). 

The evolution of phase diagrams with $Q$ opens up the possibility of tuning between different phases by applying an in-plane magnetic field. Fig.~\ref{fig:phase_varyQ} shows two phase diagrams in which the in-plane field parameter $Q$ is along one axis. Fig.~\ref{fig:phase_varyQ}(a) shows the phase diagram in the $(A,Q)$ plane at fixed $E_g=0.5\,Ry^*$. At small $A$, increasing the magnetic field turns the nematic insulator phase into the DW phase. At large $A$, the magnetic field drives the QSH state into the normal insulator state. The QAH state shows up at intermediate $A$. Fig.~\ref{fig:phase_varyQ}(b) shows the phase diagram in $(Q,E_g)$ plane at fixed $A=0.6\,a_B^* Ry^*$. At large positive $E_g$, the system stays in the normal insulator state. At small or negative $E_g$, the system starts from the QAH or QSH state and ends up in one of the three density-wave states as the magnetic field gets stronger.

\section{Discussion} \label{sec:discussion}

This paper describes a study of the influence of in-plane magnetic fields on the 
many-electron ground states of two-dimensional electron gas systems with a conduction band 
in one layer hybridized with a valence band in a nearby layer.  The interesting regime is one in which 
the spatially indirect gap is smaller than the corresponding exciton binding energy, so that 
the ground state has electron-hole coherence even when hybridization is neglected.
The role of an in-plane field is to associate a momentum boost with inter-layer tunneling processes.
The phase diagrams we construct as a function of the energy gap $E_g$ and the hybridization strength 
$A$ can in principle be tested experimentally by fabricating devices containing interfaces 
between InAs and GaSb, or other materials combinations with appropriate band lineups,
and using dual gates to tune $E_g$ at fixed electron density.  In the InAs/GaSb case
the hybridization strength $A$ can be reduced by inserting an AlSb barrier layers \cite{wu2019electrically,wu2019resistive} 
between the InAs and GaSb layers.  Our study builds on earlier work \cite{xue2018time} which studied
the influence of interactions on the phase transition between ordinary and quantum spin Hall insulators
in the absence of a magnetic field, and on work \cite{hu2017topological} which studied InAs/GaSb interfaces 
in the presence of a perpendicular field but did not identify all competing ordered phases.
Some related experimental progress has already been reported 
in recent transport experiments by Du and collaborators \cite{du2017evidence,wu2019electrically,wu2019resistive}.
Our rich theoretical phase diagrams suggest that there is much more to discover.

In a non-interacting electron theory, an in-plane magnetic field closes 
the hybridization gaps that appear for $E_g<0$ and converts the 
neutral system from insulators into semimetals.  The magnetic field strength needed to 
close the gap increases with the strength of the hybridization parameter $A$.
When interactions are included, the ground state remains insulating at all magnetic fields, by breaking 
translational symmetry to establish coherence between electron and hole states that have been boosted to different momenta.  
The observation of a gap under a strong in-plane magnetic field in experiment \cite{du2017evidence,wu2019resistive} 
is likely to be of many-body origin, and can be attributed to the density-wave states studied here. 
In the parameter range studied in this work, the semimetal state is never stable against the 
formation of density waves.  More direct evidence for density-wave states could come from transport measurements that show nonlinear current-voltage characteristics \cite{gruner1988dynamics,lee1974conductivity,fukuyama1978dynamics,lee1979electric}.

Charge density wave states are often observed experimentally in bulk three-dimensional 
narrow gap semiconductors or semimetals in which conduction band minima and valence band maxima 
occur at different wavevectors.  Recent examples include TiSe$_2$ and Ta$_2$NiSe$_5$ \cite{cercellier2007evidence,kogar2017signatures,seki2014excitonic,lu2017zero}.  There is typically 
some debate about the origin of charge density-wave states in this type of system.  We take the view that 
they can almost all be regarded as exciton-insulators \footnote{The use of the term condensate is usually taken to mean
that there is an analogy to Bose-Einstein condensation, but it should be kept in mind that this analogy is always 
imprecise since the number of electrons and holes are not conserved separately in bulk three dimensional crystals.}, in the 
same sense as the charge density wave states studied in this paper can be regarded as exciton-insulators.  
The use of this terminology to classify the type of charge density wave state is meant to suggest that if only
the band gap of the system could be varied, the ordered state would appear when the minimum 
energy of the excitonic collective modes, always present
below the interband particle-hole continuum, vanishes.  The bilayer hybridized electron-hole systems studied 
in this paper have the advantage that the key microscopic parameters of excitonic charge-density-wave systems,
the energy gap and the ordering wavevector, can indeed be varied.  Their experimental study 
therefore has the potential to draw a clear line connecting this type of charge density wave to 
ideal exciton condensates.

\begin{acknowledgements}
This work was supported by the National Science Foundation through the Center for Dynamics and Control of Materials: an NSF MRSEC under Cooperative Agreement No. DMR-1720595.
\end{acknowledgements}

\bibliography{dw_qsh}

\begin{thebibliography}{43}%
\makeatletter
\providecommand \@ifxundefined [1]{%
 \@ifx{#1\undefined}
}%
\providecommand \@ifnum [1]{%
 \ifnum #1\expandafter \@firstoftwo
 \else \expandafter \@secondoftwo
 \fi
}%
\providecommand \@ifx [1]{%
 \ifx #1\expandafter \@firstoftwo
 \else \expandafter \@secondoftwo
 \fi
}%
\providecommand \natexlab [1]{#1}%
\providecommand \enquote  [1]{``#1''}%
\providecommand \bibnamefont  [1]{#1}%
\providecommand \bibfnamefont [1]{#1}%
\providecommand \citenamefont [1]{#1}%
\providecommand \href@noop [0]{\@secondoftwo}%
\providecommand \href [0]{\begingroup \@sanitize@url \@href}%
\providecommand \@href[1]{\@@startlink{#1}\@@href}%
\providecommand \@@href[1]{\endgroup#1\@@endlink}%
\providecommand \@sanitize@url [0]{\catcode `\\12\catcode `\$12\catcode
  `\&12\catcode `\#12\catcode `\^12\catcode `\_12\catcode `\%12\relax}%
\providecommand \@@startlink[1]{}%
\providecommand \@@endlink[0]{}%
\providecommand \url  [0]{\begingroup\@sanitize@url \@url }%
\providecommand \@url [1]{\endgroup\@href {#1}{\urlprefix }}%
\providecommand \urlprefix  [0]{URL }%
\providecommand \Eprint [0]{\href }%
\providecommand \doibase [0]{https://doi.org/}%
\providecommand \selectlanguage [0]{\@gobble}%
\providecommand \bibinfo  [0]{\@secondoftwo}%
\providecommand \bibfield  [0]{\@secondoftwo}%
\providecommand \translation [1]{[#1]}%
\providecommand \BibitemOpen [0]{}%
\providecommand \bibitemStop [0]{}%
\providecommand \bibitemNoStop [0]{.\EOS\space}%
\providecommand \EOS [0]{\spacefactor3000\relax}%
\providecommand \BibitemShut  [1]{\csname bibitem#1\endcsname}%
\let\auto@bib@innerbib\@empty
\bibitem [{\citenamefont {Kane}\ and\ \citenamefont
  {Mele}(2005{\natexlab{a}})}]{KaneMele_QSH}%
  \BibitemOpen
  \bibfield  {author} {\bibinfo {author} {\bibfnamefont {C.~L.}\ \bibnamefont
  {Kane}}and\ \bibinfo {author} {\bibfnamefont {E.~J.}\ \bibnamefont {Mele}},\
  }\bibfield  {title} {\bibinfo {title} {Quantum spin {Hall} effect in
  graphene},\ }\href {https://doi.org/10.1103/PhysRevLett.95.226801} {\bibfield
   {journal} {\bibinfo  {journal} {Phys. Rev. Lett.}\ }\textbf {\bibinfo
  {volume} {95}},\ \bibinfo {pages} {226801} (\bibinfo {year}
  {2005}{\natexlab{a}})}\BibitemShut {NoStop}%
\bibitem [{\citenamefont {Kane}\ and\ \citenamefont
  {Mele}(2005{\natexlab{b}})}]{KaneMele_Z2}%
  \BibitemOpen
  \bibfield  {author} {\bibinfo {author} {\bibfnamefont {C.~L.}\ \bibnamefont
  {Kane}}and\ \bibinfo {author} {\bibfnamefont {E.~J.}\ \bibnamefont {Mele}},\
  }\bibfield  {title} {\bibinfo {title} {${Z}_{2}$ topological order and the
  quantum spin hall effect},\ }\href
  {https://doi.org/10.1103/PhysRevLett.95.146802} {\bibfield  {journal}
  {\bibinfo  {journal} {Phys. Rev. Lett.}\ }\textbf {\bibinfo {volume} {95}},\
  \bibinfo {pages} {146802} (\bibinfo {year} {2005}{\natexlab{b}})}\BibitemShut
  {NoStop}%
\bibitem [{\citenamefont {Bernevig}\ \emph {et~al.}(2006)\citenamefont
  {Bernevig}, \citenamefont {Hughes},\ and\ \citenamefont
  {Zhang}}]{bernevig2006quantum}%
  \BibitemOpen
  \bibfield  {author} {\bibinfo {author} {\bibfnamefont {B.~A.}\ \bibnamefont
  {Bernevig}}, \bibinfo {author} {\bibfnamefont {T.~L.}\ \bibnamefont
  {Hughes}}, and\ \bibinfo {author} {\bibfnamefont {S.-C.}\ \bibnamefont
  {Zhang}},\ }\bibfield  {title} {\bibinfo {title} {Quantum spin \rm{H}all
  effect and topological phase transition in \rm{HgTe} quantum wells},\
  }\href@noop {} {\bibfield  {journal} {\bibinfo  {journal} {Science}\ }\textbf
  {\bibinfo {volume} {314}},\ \bibinfo {pages} {1757} (\bibinfo {year}
  {2006})}\BibitemShut {NoStop}%
\bibitem [{\citenamefont {K{\"o}nig}\ \emph {et~al.}(2007)\citenamefont
  {K{\"o}nig}, \citenamefont {Wiedmann}, \citenamefont {Br{\"u}ne},
  \citenamefont {Roth}, \citenamefont {Buhmann}, \citenamefont {Molenkamp},
  \citenamefont {Qi},\ and\ \citenamefont {Zhang}}]{konig2007quantum}%
  \BibitemOpen
  \bibfield  {author} {\bibinfo {author} {\bibfnamefont {M.}~\bibnamefont
  {K{\"o}nig}}, \bibinfo {author} {\bibfnamefont {S.}~\bibnamefont {Wiedmann}},
  \bibinfo {author} {\bibfnamefont {C.}~\bibnamefont {Br{\"u}ne}}, \bibinfo
  {author} {\bibfnamefont {A.}~\bibnamefont {Roth}}, \bibinfo {author}
  {\bibfnamefont {H.}~\bibnamefont {Buhmann}}, \bibinfo {author} {\bibfnamefont
  {L.~W.}\ \bibnamefont {Molenkamp}}, \bibinfo {author} {\bibfnamefont {X.-L.}\
  \bibnamefont {Qi}}, and\ \bibinfo {author} {\bibfnamefont {S.-C.}\
  \bibnamefont {Zhang}},\ }\bibfield  {title} {\bibinfo {title} {\rm{Quantum
  spin Hall insulator state in HgTe quantum wells}},\ }\href@noop {} {\bibfield
   {journal} {\bibinfo  {journal} {Science}\ }\textbf {\bibinfo {volume}
  {318}},\ \bibinfo {pages} {766} (\bibinfo {year} {2007})}\BibitemShut
  {NoStop}%
\bibitem [{\citenamefont {Liu}\ \emph {et~al.}(2008)\citenamefont {Liu},
  \citenamefont {Hughes}, \citenamefont {Qi}, \citenamefont {Wang},\ and\
  \citenamefont {Zhang}}]{liu2008quantum}%
  \BibitemOpen
  \bibfield  {author} {\bibinfo {author} {\bibfnamefont {C.}~\bibnamefont
  {Liu}}, \bibinfo {author} {\bibfnamefont {T.~L.}\ \bibnamefont {Hughes}},
  \bibinfo {author} {\bibfnamefont {X.-L.}\ \bibnamefont {Qi}}, \bibinfo
  {author} {\bibfnamefont {K.}~\bibnamefont {Wang}}, and\ \bibinfo {author}
  {\bibfnamefont {S.-C.}\ \bibnamefont {Zhang}},\ }\bibfield  {title} {\bibinfo
  {title} {\rm{Quantum spin Hall effect in inverted type-II semiconductors}},\
  }\href@noop {} {\bibfield  {journal} {\bibinfo  {journal} {Physical review
  letters}\ }\textbf {\bibinfo {volume} {100}},\ \bibinfo {pages} {236601}
  (\bibinfo {year} {2008})}\BibitemShut {NoStop}%
\bibitem [{\citenamefont {Knez}\ \emph {et~al.}(2011)\citenamefont {Knez},
  \citenamefont {Du},\ and\ \citenamefont {Sullivan}}]{knez2011evidence}%
  \BibitemOpen
  \bibfield  {author} {\bibinfo {author} {\bibfnamefont {I.}~\bibnamefont
  {Knez}}, \bibinfo {author} {\bibfnamefont {R.-R.}\ \bibnamefont {Du}}, and\
  \bibinfo {author} {\bibfnamefont {G.}~\bibnamefont {Sullivan}},\ }\bibfield
  {title} {\bibinfo {title} {Evidence for helical edge modes in inverted
  \rm{InAs/GaSb} quantum wells},\ }\href@noop {} {\bibfield  {journal}
  {\bibinfo  {journal} {Physical review letters}\ }\textbf {\bibinfo {volume}
  {107}},\ \bibinfo {pages} {136603} (\bibinfo {year} {2011})}\BibitemShut
  {NoStop}%
\bibitem [{\citenamefont {Du}\ \emph {et~al.}(2015)\citenamefont {Du},
  \citenamefont {Knez}, \citenamefont {Sullivan},\ and\ \citenamefont
  {Du}}]{du2015robust}%
  \BibitemOpen
  \bibfield  {author} {\bibinfo {author} {\bibfnamefont {L.}~\bibnamefont
  {Du}}, \bibinfo {author} {\bibfnamefont {I.}~\bibnamefont {Knez}}, \bibinfo
  {author} {\bibfnamefont {G.}~\bibnamefont {Sullivan}}, and\ \bibinfo {author}
  {\bibfnamefont {R.-R.}\ \bibnamefont {Du}},\ }\bibfield  {title} {\bibinfo
  {title} {Robust helical edge transport in gated inas/gasb bilayers},\
  }\href@noop {} {\bibfield  {journal} {\bibinfo  {journal} {Physical review
  letters}\ }\textbf {\bibinfo {volume} {114}},\ \bibinfo {pages} {096802}
  (\bibinfo {year} {2015})}\BibitemShut {NoStop}%
\bibitem [{\citenamefont {Keldysh}\ and\ \citenamefont
  {Kopaev}(1965)}]{Keldysh1965}%
  \BibitemOpen
  \bibfield  {author} {\bibinfo {author} {\bibfnamefont {L.~V.}\ \bibnamefont
  {Keldysh}}and\ \bibinfo {author} {\bibfnamefont {Y.~V.}\ \bibnamefont
  {Kopaev}},\ }\bibfield  {title} {\bibinfo {title} {Possible instability of
  semimetallic state toward coulomb interaction},\ }\href@noop {} {\bibfield
  {journal} {\bibinfo  {journal} {Sov. Phys. Solid State}\ }\textbf {\bibinfo
  {volume} {6}},\ \bibinfo {pages} {2219} (\bibinfo {year} {1965})}\BibitemShut
  {NoStop}%
\bibitem [{\citenamefont {J{\'e}rome}\ \emph {et~al.}(1967)\citenamefont
  {J{\'e}rome}, \citenamefont {Rice},\ and\ \citenamefont
  {Kohn}}]{jerome1967excitonic}%
  \BibitemOpen
  \bibfield  {author} {\bibinfo {author} {\bibfnamefont {D.}~\bibnamefont
  {J{\'e}rome}}, \bibinfo {author} {\bibfnamefont {T.}~\bibnamefont {Rice}},
  and\ \bibinfo {author} {\bibfnamefont {W.}~\bibnamefont {Kohn}},\ }\bibfield
  {title} {\bibinfo {title} {Excitonic insulator},\ }\href@noop {} {\bibfield
  {journal} {\bibinfo  {journal} {Physical Review}\ }\textbf {\bibinfo {volume}
  {158}},\ \bibinfo {pages} {462} (\bibinfo {year} {1967})}\BibitemShut
  {NoStop}%
\bibitem [{\citenamefont {Lozovik}\ and\ \citenamefont
  {Yudson}(1976)}]{Lozovik1976}%
  \BibitemOpen
  \bibfield  {author} {\bibinfo {author} {\bibfnamefont {Y.~E.}\ \bibnamefont
  {Lozovik}}and\ \bibinfo {author} {\bibfnamefont {V.~I.}\ \bibnamefont
  {Yudson}},\ }\bibfield  {title} {\bibinfo {title} {A new mechanism for
  superconductivity: pairing between spatially separated electrons and holes},\
  }\href@noop {} {\bibfield  {journal} {\bibinfo  {journal} {Sov. Phys. JETP}\
  }\textbf {\bibinfo {volume} {44}},\ \bibinfo {pages} {389} (\bibinfo {year}
  {1976})}\BibitemShut {NoStop}%
\bibitem [{\citenamefont {Comte}\ and\ \citenamefont
  {Nozieres}(1982)}]{Comte1982}%
  \BibitemOpen
  \bibfield  {author} {\bibinfo {author} {\bibfnamefont {C.}~\bibnamefont
  {Comte}}and\ \bibinfo {author} {\bibfnamefont {P.}~\bibnamefont {Nozieres}},\
  }\bibfield  {title} {\bibinfo {title} {{Exciton Bose condensation: the ground
  state of an electron-hole gas-I. Mean field description of a simplified
  model}},\ }\href@noop {} {\bibfield  {journal} {\bibinfo  {journal} {J. Phys.
  (Paris)}\ }\textbf {\bibinfo {volume} {43}},\ \bibinfo {pages} {1069}
  (\bibinfo {year} {1982})}\BibitemShut {NoStop}%
\bibitem [{\citenamefont {Zhu}\ \emph {et~al.}(1995)\citenamefont {Zhu},
  \citenamefont {Littlewood}, \citenamefont {Hybertsen},\ and\ \citenamefont
  {Rice}}]{zhu1995exciton}%
  \BibitemOpen
  \bibfield  {author} {\bibinfo {author} {\bibfnamefont {X.}~\bibnamefont
  {Zhu}}, \bibinfo {author} {\bibfnamefont {P.}~\bibnamefont {Littlewood}},
  \bibinfo {author} {\bibfnamefont {M.~S.}\ \bibnamefont {Hybertsen}}, and\
  \bibinfo {author} {\bibfnamefont {T.}~\bibnamefont {Rice}},\ }\bibfield
  {title} {\bibinfo {title} {Exciton condensate in semiconductor quantum well
  structures},\ }\href@noop {} {\bibfield  {journal} {\bibinfo  {journal}
  {Physical review letters}\ }\textbf {\bibinfo {volume} {74}},\ \bibinfo
  {pages} {1633} (\bibinfo {year} {1995})}\BibitemShut {NoStop}%
\bibitem [{\citenamefont {Lozovik}\ and\ \citenamefont
  {Berman}(1996)}]{lozovik1996phase}%
  \BibitemOpen
  \bibfield  {author} {\bibinfo {author} {\bibfnamefont {Y.~E.}\ \bibnamefont
  {Lozovik}}and\ \bibinfo {author} {\bibfnamefont {O.}~\bibnamefont {Berman}},\
  }\bibfield  {title} {\bibinfo {title} {Phase transitions in a system of two
  coupled quantum wells},\ }\href@noop {} {\bibfield  {journal} {\bibinfo
  {journal} {Journal of Experimental and Theoretical Physics Letters}\ }\textbf
  {\bibinfo {volume} {64}},\ \bibinfo {pages} {573} (\bibinfo {year}
  {1996})}\BibitemShut {NoStop}%
\bibitem [{\citenamefont {High}\ \emph {et~al.}(2012)\citenamefont {High},
  \citenamefont {Leonard}, \citenamefont {Hammack}, \citenamefont {Fogler},
  \citenamefont {Butov}, \citenamefont {Kavokin}, \citenamefont {Campman},\
  and\ \citenamefont {Gossard}}]{high2012spontaneous}%
  \BibitemOpen
  \bibfield  {author} {\bibinfo {author} {\bibfnamefont {A.~A.}\ \bibnamefont
  {High}}, \bibinfo {author} {\bibfnamefont {J.~R.}\ \bibnamefont {Leonard}},
  \bibinfo {author} {\bibfnamefont {A.~T.}\ \bibnamefont {Hammack}}, \bibinfo
  {author} {\bibfnamefont {M.~M.}\ \bibnamefont {Fogler}}, \bibinfo {author}
  {\bibfnamefont {L.~V.}\ \bibnamefont {Butov}}, \bibinfo {author}
  {\bibfnamefont {A.~V.}\ \bibnamefont {Kavokin}}, \bibinfo {author}
  {\bibfnamefont {K.~L.}\ \bibnamefont {Campman}}, and\ \bibinfo {author}
  {\bibfnamefont {A.~C.}\ \bibnamefont {Gossard}},\ }\bibfield  {title}
  {\bibinfo {title} {Spontaneous coherence in a cold exciton gas},\ }\href@noop
  {} {\bibfield  {journal} {\bibinfo  {journal} {Nature}\ }\textbf {\bibinfo
  {volume} {483}},\ \bibinfo {pages} {584} (\bibinfo {year}
  {2012})}\BibitemShut {NoStop}%
\bibitem [{\citenamefont {Wu}\ \emph {et~al.}(2015)\citenamefont {Wu},
  \citenamefont {Xue},\ and\ \citenamefont {MacDonald}}]{wu2015theory}%
  \BibitemOpen
  \bibfield  {author} {\bibinfo {author} {\bibfnamefont {F.-C.}\ \bibnamefont
  {Wu}}, \bibinfo {author} {\bibfnamefont {F.}~\bibnamefont {Xue}}, and\
  \bibinfo {author} {\bibfnamefont {A.}~\bibnamefont {MacDonald}},\ }\bibfield
  {title} {\bibinfo {title} {Theory of two-dimensional spatially indirect
  equilibrium exciton condensates},\ }\href@noop {} {\bibfield  {journal}
  {\bibinfo  {journal} {Physical Review B}\ }\textbf {\bibinfo {volume} {92}},\
  \bibinfo {pages} {165121} (\bibinfo {year} {2015})}\BibitemShut {NoStop}%
\bibitem [{\citenamefont {Du}\ \emph {et~al.}(2017)\citenamefont {Du},
  \citenamefont {Li}, \citenamefont {Lou}, \citenamefont {Sullivan},
  \citenamefont {Chang}, \citenamefont {Kono},\ and\ \citenamefont
  {Du}}]{du2017evidence}%
  \BibitemOpen
  \bibfield  {author} {\bibinfo {author} {\bibfnamefont {L.}~\bibnamefont
  {Du}}, \bibinfo {author} {\bibfnamefont {X.}~\bibnamefont {Li}}, \bibinfo
  {author} {\bibfnamefont {W.}~\bibnamefont {Lou}}, \bibinfo {author}
  {\bibfnamefont {G.}~\bibnamefont {Sullivan}}, \bibinfo {author}
  {\bibfnamefont {K.}~\bibnamefont {Chang}}, \bibinfo {author} {\bibfnamefont
  {J.}~\bibnamefont {Kono}}, and\ \bibinfo {author} {\bibfnamefont {R.-R.}\
  \bibnamefont {Du}},\ }\bibfield  {title} {\bibinfo {title} {Evidence for a
  topological excitonic insulator in \rm{InAs/GaSb} bilayers},\ }\href@noop {}
  {\bibfield  {journal} {\bibinfo  {journal} {Nature communications}\ }\textbf
  {\bibinfo {volume} {8}},\ \bibinfo {pages} {1} (\bibinfo {year}
  {2017})}\BibitemShut {NoStop}%
\bibitem [{\citenamefont {Wu}\ \emph {et~al.}(2019{\natexlab{a}})\citenamefont
  {Wu}, \citenamefont {Lou}, \citenamefont {Chang}, \citenamefont {Sullivan},
  \citenamefont {Ikhlassi},\ and\ \citenamefont {Du}}]{wu2019electrically}%
  \BibitemOpen
  \bibfield  {author} {\bibinfo {author} {\bibfnamefont {X.-J.}\ \bibnamefont
  {Wu}}, \bibinfo {author} {\bibfnamefont {W.}~\bibnamefont {Lou}}, \bibinfo
  {author} {\bibfnamefont {K.}~\bibnamefont {Chang}}, \bibinfo {author}
  {\bibfnamefont {G.}~\bibnamefont {Sullivan}}, \bibinfo {author}
  {\bibfnamefont {A.}~\bibnamefont {Ikhlassi}}, and\ \bibinfo {author}
  {\bibfnamefont {R.-R.}\ \bibnamefont {Du}},\ }\bibfield  {title} {\bibinfo
  {title} {Electrically tuning many-body states in a coulomb-coupled
  \rm{InAs/InGaSb} double layer},\ }\href@noop {} {\bibfield  {journal}
  {\bibinfo  {journal} {Physical Review B}\ }\textbf {\bibinfo {volume}
  {100}},\ \bibinfo {pages} {165309} (\bibinfo {year}
  {2019}{\natexlab{a}})}\BibitemShut {NoStop}%
\bibitem [{\citenamefont {Wu}\ \emph {et~al.}(2019{\natexlab{b}})\citenamefont
  {Wu}, \citenamefont {Lou}, \citenamefont {Chang}, \citenamefont {Sullivan},\
  and\ \citenamefont {Du}}]{wu2019resistive}%
  \BibitemOpen
  \bibfield  {author} {\bibinfo {author} {\bibfnamefont {X.}~\bibnamefont
  {Wu}}, \bibinfo {author} {\bibfnamefont {W.}~\bibnamefont {Lou}}, \bibinfo
  {author} {\bibfnamefont {K.}~\bibnamefont {Chang}}, \bibinfo {author}
  {\bibfnamefont {G.}~\bibnamefont {Sullivan}}, and\ \bibinfo {author}
  {\bibfnamefont {R.-R.}\ \bibnamefont {Du}},\ }\bibfield  {title} {\bibinfo
  {title} {Resistive signature of excitonic coupling in an electron-hole double
  layer with a middle barrier},\ }\href@noop {} {\bibfield  {journal} {\bibinfo
   {journal} {Physical Review B}\ }\textbf {\bibinfo {volume} {99}},\ \bibinfo
  {pages} {085307} (\bibinfo {year} {2019}{\natexlab{b}})}\BibitemShut
  {NoStop}%
\bibitem [{\citenamefont {Budich}\ \emph {et~al.}(2014)\citenamefont {Budich},
  \citenamefont {Trauzettel},\ and\ \citenamefont {Michetti}}]{budich2014time}%
  \BibitemOpen
  \bibfield  {author} {\bibinfo {author} {\bibfnamefont {J.~C.}\ \bibnamefont
  {Budich}}, \bibinfo {author} {\bibfnamefont {B.}~\bibnamefont {Trauzettel}},
  and\ \bibinfo {author} {\bibfnamefont {P.}~\bibnamefont {Michetti}},\
  }\bibfield  {title} {\bibinfo {title} {Time reversal symmetric topological
  exciton condensate in bilayer \rm{HgTe} quantum wells},\ }\href@noop {}
  {\bibfield  {journal} {\bibinfo  {journal} {Physical review letters}\
  }\textbf {\bibinfo {volume} {112}},\ \bibinfo {pages} {146405} (\bibinfo
  {year} {2014})}\BibitemShut {NoStop}%
\bibitem [{\citenamefont {Pikulin}\ and\ \citenamefont
  {Hyart}(2014)}]{pikulin2014interplay}%
  \BibitemOpen
  \bibfield  {author} {\bibinfo {author} {\bibfnamefont {D.}~\bibnamefont
  {Pikulin}}and\ \bibinfo {author} {\bibfnamefont {T.}~\bibnamefont {Hyart}},\
  }\bibfield  {title} {\bibinfo {title} {Interplay of exciton condensation and
  the quantum spin \rm{H}all effect in \rm{InAs/GaSb} bilayers},\ }\href@noop
  {} {\bibfield  {journal} {\bibinfo  {journal} {Physical Review Letters}\
  }\textbf {\bibinfo {volume} {112}},\ \bibinfo {pages} {176403} (\bibinfo
  {year} {2014})}\BibitemShut {NoStop}%
\bibitem [{\citenamefont {Hu}\ \emph {et~al.}(2017)\citenamefont {Hu},
  \citenamefont {Chen}, \citenamefont {Liu}, \citenamefont {Zhang},\ and\
  \citenamefont {Zhou}}]{hu2017topological}%
  \BibitemOpen
  \bibfield  {author} {\bibinfo {author} {\bibfnamefont {L.-H.}\ \bibnamefont
  {Hu}}, \bibinfo {author} {\bibfnamefont {C.-C.}\ \bibnamefont {Chen}},
  \bibinfo {author} {\bibfnamefont {C.-X.}\ \bibnamefont {Liu}}, \bibinfo
  {author} {\bibfnamefont {F.-C.}\ \bibnamefont {Zhang}}, and\ \bibinfo
  {author} {\bibfnamefont {Y.}~\bibnamefont {Zhou}},\ }\bibfield  {title}
  {\bibinfo {title} {Topological charge-density and spin-density waves in
  \rm{InAs/GaSb} quantum wells under an in-plane magnetic field},\ }\href@noop
  {} {\bibfield  {journal} {\bibinfo  {journal} {Physical Review B}\ }\textbf
  {\bibinfo {volume} {96}},\ \bibinfo {pages} {075130} (\bibinfo {year}
  {2017})}\BibitemShut {NoStop}%
\bibitem [{\citenamefont {Xue}\ and\ \citenamefont
  {MacDonald}(2018)}]{xue2018time}%
  \BibitemOpen
  \bibfield  {author} {\bibinfo {author} {\bibfnamefont {F.}~\bibnamefont
  {Xue}}and\ \bibinfo {author} {\bibfnamefont {A.~H.}\ \bibnamefont
  {MacDonald}},\ }\bibfield  {title} {\bibinfo {title} {Time-reversal
  symmetry-breaking nematic insulators near quantum spin \rm{H}all phase
  transitions},\ }\href@noop {} {\bibfield  {journal} {\bibinfo  {journal}
  {Physical Review Letters}\ }\textbf {\bibinfo {volume} {120}},\ \bibinfo
  {pages} {186802} (\bibinfo {year} {2018})}\BibitemShut {NoStop}%
\bibitem [{\citenamefont {Hu}\ \emph {et~al.}(2016)\citenamefont {Hu},
  \citenamefont {Xu}, \citenamefont {Zhang},\ and\ \citenamefont
  {Zhou}}]{Hu2016}%
  \BibitemOpen
  \bibfield  {author} {\bibinfo {author} {\bibfnamefont {L.-H.}\ \bibnamefont
  {Hu}}, \bibinfo {author} {\bibfnamefont {D.-H.}\ \bibnamefont {Xu}}, \bibinfo
  {author} {\bibfnamefont {F.-C.}\ \bibnamefont {Zhang}}, and\ \bibinfo
  {author} {\bibfnamefont {Y.}~\bibnamefont {Zhou}},\ }\bibfield  {title}
  {\bibinfo {title} {Effect of in-plane magnetic field and applied strain in
  quantum spin hall systems: Application to {InAs/GaSb} quantum wells},\ }\href
  {https://doi.org/10.1103/PhysRevB.94.085306} {\bibfield  {journal} {\bibinfo
  {journal} {Phys. Rev. B}\ }\textbf {\bibinfo {volume} {94}},\ \bibinfo
  {pages} {085306} (\bibinfo {year} {2016})}\BibitemShut {NoStop}%
\bibitem [{\citenamefont {Levinshtein}\ \emph {et~al.}(1996)\citenamefont
  {Levinshtein}, \citenamefont {Rumyantsev},\ and\ \citenamefont
  {Shur}}]{levinshtein1996handbook}%
  \BibitemOpen
  \bibfield  {author} {\bibinfo {author} {\bibfnamefont {M.}~\bibnamefont
  {Levinshtein}}, \bibinfo {author} {\bibfnamefont {S.}~\bibnamefont
  {Rumyantsev}}, and\ \bibinfo {author} {\bibfnamefont {M.}~\bibnamefont
  {Shur}},\ }\href {https://doi.org/10.1142/2046-vol1} {\emph {\bibinfo {title}
  {Handbook Series on Semiconductor Parameters}}}\ (\bibinfo  {publisher}
  {WORLD SCIENTIFIC},\ \bibinfo {year} {1996})\BibitemShut {NoStop}%
\bibitem [{\citenamefont {Liu}\ and\ \citenamefont
  {Zhang}(2013)}]{liu2013models}%
  \BibitemOpen
  \bibfield  {author} {\bibinfo {author} {\bibfnamefont {C.}~\bibnamefont
  {Liu}}and\ \bibinfo {author} {\bibfnamefont {S.}~\bibnamefont {Zhang}},\
  }\bibfield  {title} {\bibinfo {title} {Models and materials for topological
  insulators},\ }in\ \href@noop {} {\emph {\bibinfo {booktitle} {Contemporary
  Concepts of Condensed Matter Science}}},\ Vol.~\bibinfo {volume} {6}\
  (\bibinfo  {publisher} {Elsevier},\ \bibinfo {year} {2013})\ pp.\ \bibinfo
  {pages} {59--89}\BibitemShut {NoStop}%
\bibitem [{Note1()}]{Note1}%
  \BibitemOpen
  \bibinfo {note} {In this summary, we have ignored some minor phases that
  occur in a small region of the phase diagram near the $A=0$
  line.}\BibitemShut {Stop}%
\bibitem [{\citenamefont {Sheng}\ \emph {et~al.}(2006)\citenamefont {Sheng},
  \citenamefont {Weng}, \citenamefont {Sheng},\ and\ \citenamefont
  {Haldane}}]{sheng2006quantum}%
  \BibitemOpen
  \bibfield  {author} {\bibinfo {author} {\bibfnamefont {D.}~\bibnamefont
  {Sheng}}, \bibinfo {author} {\bibfnamefont {Z.}~\bibnamefont {Weng}},
  \bibinfo {author} {\bibfnamefont {L.}~\bibnamefont {Sheng}}, and\ \bibinfo
  {author} {\bibfnamefont {F.}~\bibnamefont {Haldane}},\ }\bibfield  {title}
  {\bibinfo {title} {Quantum spin-\rm{H}all effect and topologically invariant
  \rm{C}hern numbers},\ }\href@noop {} {\bibfield  {journal} {\bibinfo
  {journal} {Physical Review Letters}\ }\textbf {\bibinfo {volume} {97}},\
  \bibinfo {pages} {036808} (\bibinfo {year} {2006})}\BibitemShut {NoStop}%
\bibitem [{\citenamefont {Prodan}(2009)}]{prodan2009robustness}%
  \BibitemOpen
  \bibfield  {author} {\bibinfo {author} {\bibfnamefont {E.}~\bibnamefont
  {Prodan}},\ }\bibfield  {title} {\bibinfo {title} {Robustness of the
  spin-\rm{C}hern number},\ }\href@noop {} {\bibfield  {journal} {\bibinfo
  {journal} {Physical Review B}\ }\textbf {\bibinfo {volume} {80}},\ \bibinfo
  {pages} {125327} (\bibinfo {year} {2009})}\BibitemShut {NoStop}%
\bibitem [{\citenamefont {Yang}\ \emph {et~al.}(2011)\citenamefont {Yang},
  \citenamefont {Xu}, \citenamefont {Sheng}, \citenamefont {Wang},
  \citenamefont {Xing},\ and\ \citenamefont {Sheng}}]{yang2011time}%
  \BibitemOpen
  \bibfield  {author} {\bibinfo {author} {\bibfnamefont {Y.}~\bibnamefont
  {Yang}}, \bibinfo {author} {\bibfnamefont {Z.}~\bibnamefont {Xu}}, \bibinfo
  {author} {\bibfnamefont {L.}~\bibnamefont {Sheng}}, \bibinfo {author}
  {\bibfnamefont {B.}~\bibnamefont {Wang}}, \bibinfo {author} {\bibfnamefont
  {D.}~\bibnamefont {Xing}}, and\ \bibinfo {author} {\bibfnamefont
  {D.}~\bibnamefont {Sheng}},\ }\bibfield  {title} {\bibinfo {title}
  {Time-reversal-symmetry-broken quantum spin \rm{H}all effect},\ }\href@noop
  {} {\bibfield  {journal} {\bibinfo  {journal} {Physical Review Letters}\
  }\textbf {\bibinfo {volume} {107}},\ \bibinfo {pages} {066602} (\bibinfo
  {year} {2011})}\BibitemShut {NoStop}%
\bibitem [{\citenamefont {Vanderbilt}(2018)}]{vanderbilt2018berry}%
  \BibitemOpen
  \bibfield  {author} {\bibinfo {author} {\bibfnamefont {D.}~\bibnamefont
  {Vanderbilt}},\ }\href@noop {} {\emph {\bibinfo {title} {Berry Phases in
  Electronic Structure Theory: Electric Polarization, Orbital Magnetization and
  Topological Insulators}}}\ (\bibinfo  {publisher} {Cambridge University
  Press},\ \bibinfo {year} {2018})\ p.\ \bibinfo {pages} {226}\BibitemShut
  {NoStop}%
\bibitem [{Note2()}]{Note2}%
  \BibitemOpen
  \bibinfo {note} {We have ignored some minor phases that appear near the phase
  boundaries. We find that the QSH/DW and QAH/DW phases are more stable as the
  magnetic field strength increases.}\BibitemShut {Stop}%
\bibitem [{\citenamefont {Yu}\ \emph {et~al.}(2011)\citenamefont {Yu},
  \citenamefont {Qi}, \citenamefont {Bernevig}, \citenamefont {Fang},\ and\
  \citenamefont {Dai}}]{yu2011equivalent}%
  \BibitemOpen
  \bibfield  {author} {\bibinfo {author} {\bibfnamefont {R.}~\bibnamefont
  {Yu}}, \bibinfo {author} {\bibfnamefont {X.~L.}\ \bibnamefont {Qi}}, \bibinfo
  {author} {\bibfnamefont {A.}~\bibnamefont {Bernevig}}, \bibinfo {author}
  {\bibfnamefont {Z.}~\bibnamefont {Fang}}, and\ \bibinfo {author}
  {\bibfnamefont {X.}~\bibnamefont {Dai}},\ }\bibfield  {title} {\bibinfo
  {title} {Equivalent expression of $\mathbb{Z}_2$ topological invariant for
  band insulators using the non-\rm{Abelian Berry} connection},\ }\href@noop {}
  {\bibfield  {journal} {\bibinfo  {journal} {Physical Review B}\ }\textbf
  {\bibinfo {volume} {84}},\ \bibinfo {pages} {075119} (\bibinfo {year}
  {2011})}\BibitemShut {NoStop}%
\bibitem [{\citenamefont {Weng}\ \emph {et~al.}(2015)\citenamefont {Weng},
  \citenamefont {Yu}, \citenamefont {Hu}, \citenamefont {Dai},\ and\
  \citenamefont {Fang}}]{weng2015quantum}%
  \BibitemOpen
  \bibfield  {author} {\bibinfo {author} {\bibfnamefont {H.}~\bibnamefont
  {Weng}}, \bibinfo {author} {\bibfnamefont {R.}~\bibnamefont {Yu}}, \bibinfo
  {author} {\bibfnamefont {X.}~\bibnamefont {Hu}}, \bibinfo {author}
  {\bibfnamefont {X.}~\bibnamefont {Dai}}, and\ \bibinfo {author}
  {\bibfnamefont {Z.}~\bibnamefont {Fang}},\ }\bibfield  {title} {\bibinfo
  {title} {Quantum anomalous \rm{H}all effect and related topological
  electronic states},\ }\href@noop {} {\bibfield  {journal} {\bibinfo
  {journal} {Advances in Physics}\ }\textbf {\bibinfo {volume} {64}},\ \bibinfo
  {pages} {227} (\bibinfo {year} {2015})}\BibitemShut {NoStop}%
\bibitem [{Note3()}]{Note3}%
  \BibitemOpen
  \bibinfo {note} {In principle we may apply the definition of spin Chern
  number in Ref.~\protect \rev@citealp {prodan2009robustness}. However, here we
  follow a more conservative convention and do not calculate the spin Chern
  number for the QSH/DW state, because the definition involves a rather
  artificial requirement on adiabatic path \cite {vanderbilt2018berry} and
  lacks physical significance.}\BibitemShut {Stop}%
\bibitem [{\citenamefont {Gr{\"u}ner}(1988)}]{gruner1988dynamics}%
  \BibitemOpen
  \bibfield  {author} {\bibinfo {author} {\bibfnamefont {G.}~\bibnamefont
  {Gr{\"u}ner}},\ }\bibfield  {title} {\bibinfo {title} {The dynamics of
  charge-density waves},\ }\href@noop {} {\bibfield  {journal} {\bibinfo
  {journal} {Reviews of modern physics}\ }\textbf {\bibinfo {volume} {60}},\
  \bibinfo {pages} {1129} (\bibinfo {year} {1988})}\BibitemShut {NoStop}%
\bibitem [{\citenamefont {Lee}\ \emph {et~al.}(1974)\citenamefont {Lee},
  \citenamefont {Rice},\ and\ \citenamefont {Anderson}}]{lee1974conductivity}%
  \BibitemOpen
  \bibfield  {author} {\bibinfo {author} {\bibfnamefont {P.}~\bibnamefont
  {Lee}}, \bibinfo {author} {\bibfnamefont {T.}~\bibnamefont {Rice}}, and\
  \bibinfo {author} {\bibfnamefont {P.}~\bibnamefont {Anderson}},\ }\bibfield
  {title} {\bibinfo {title} {Conductivity from charge or spin density waves},\
  }\href@noop {} {\bibfield  {journal} {\bibinfo  {journal} {Solid State
  Communications}\ }\textbf {\bibinfo {volume} {14}},\ \bibinfo {pages} {703}
  (\bibinfo {year} {1974})}\BibitemShut {NoStop}%
\bibitem [{\citenamefont {Fukuyama}\ and\ \citenamefont
  {Lee}(1978)}]{fukuyama1978dynamics}%
  \BibitemOpen
  \bibfield  {author} {\bibinfo {author} {\bibfnamefont {H.}~\bibnamefont
  {Fukuyama}}and\ \bibinfo {author} {\bibfnamefont {P.~A.}\ \bibnamefont
  {Lee}},\ }\bibfield  {title} {\bibinfo {title} {Dynamics of the
  charge-density wave. \rm{I. I}mpurity pinning in a single chain},\
  }\href@noop {} {\bibfield  {journal} {\bibinfo  {journal} {Physical Review
  B}\ }\textbf {\bibinfo {volume} {17}},\ \bibinfo {pages} {535} (\bibinfo
  {year} {1978})}\BibitemShut {NoStop}%
\bibitem [{\citenamefont {Lee}\ and\ \citenamefont
  {Rice}(1979)}]{lee1979electric}%
  \BibitemOpen
  \bibfield  {author} {\bibinfo {author} {\bibfnamefont {P.}~\bibnamefont
  {Lee}}and\ \bibinfo {author} {\bibfnamefont {T.}~\bibnamefont {Rice}},\
  }\bibfield  {title} {\bibinfo {title} {Electric field depinning of charge
  density waves},\ }\href@noop {} {\bibfield  {journal} {\bibinfo  {journal}
  {Physical Review B}\ }\textbf {\bibinfo {volume} {19}},\ \bibinfo {pages}
  {3970} (\bibinfo {year} {1979})}\BibitemShut {NoStop}%
\bibitem [{\citenamefont {Cercellier}\ \emph {et~al.}(2007)\citenamefont
  {Cercellier}, \citenamefont {Monney}, \citenamefont {Clerc}, \citenamefont
  {Battaglia}, \citenamefont {Despont}, \citenamefont {Garnier}, \citenamefont
  {Beck}, \citenamefont {Aebi}, \citenamefont {Patthey}, \citenamefont {Berger}
  \emph {et~al.}}]{cercellier2007evidence}%
  \BibitemOpen
  \bibfield  {author} {\bibinfo {author} {\bibfnamefont {H.}~\bibnamefont
  {Cercellier}}, \bibinfo {author} {\bibfnamefont {C.}~\bibnamefont {Monney}},
  \bibinfo {author} {\bibfnamefont {F.}~\bibnamefont {Clerc}}, \bibinfo
  {author} {\bibfnamefont {C.}~\bibnamefont {Battaglia}}, \bibinfo {author}
  {\bibfnamefont {L.}~\bibnamefont {Despont}}, \bibinfo {author} {\bibfnamefont
  {M.}~\bibnamefont {Garnier}}, \bibinfo {author} {\bibfnamefont
  {H.}~\bibnamefont {Beck}}, \bibinfo {author} {\bibfnamefont {P.}~\bibnamefont
  {Aebi}}, \bibinfo {author} {\bibfnamefont {L.}~\bibnamefont {Patthey}},
  \bibinfo {author} {\bibfnamefont {H.}~\bibnamefont {Berger}},  \emph
  {et~al.},\ }\bibfield  {title} {\bibinfo {title} {Evidence for an excitonic
  insulator phase in {1T-TiSe$_2$}},\ }\href@noop {} {\bibfield  {journal}
  {\bibinfo  {journal} {Physical review letters}\ }\textbf {\bibinfo {volume}
  {99}},\ \bibinfo {pages} {146403} (\bibinfo {year} {2007})}\BibitemShut
  {NoStop}%
\bibitem [{\citenamefont {Kogar}\ \emph {et~al.}(2017)\citenamefont {Kogar},
  \citenamefont {Rak}, \citenamefont {Vig}, \citenamefont {Husain},
  \citenamefont {Flicker}, \citenamefont {Joe}, \citenamefont {Venema},
  \citenamefont {MacDougall}, \citenamefont {Chiang}, \citenamefont {Fradkin}
  \emph {et~al.}}]{kogar2017signatures}%
  \BibitemOpen
  \bibfield  {author} {\bibinfo {author} {\bibfnamefont {A.}~\bibnamefont
  {Kogar}}, \bibinfo {author} {\bibfnamefont {M.~S.}\ \bibnamefont {Rak}},
  \bibinfo {author} {\bibfnamefont {S.}~\bibnamefont {Vig}}, \bibinfo {author}
  {\bibfnamefont {A.~A.}\ \bibnamefont {Husain}}, \bibinfo {author}
  {\bibfnamefont {F.}~\bibnamefont {Flicker}}, \bibinfo {author} {\bibfnamefont
  {Y.~I.}\ \bibnamefont {Joe}}, \bibinfo {author} {\bibfnamefont
  {L.}~\bibnamefont {Venema}}, \bibinfo {author} {\bibfnamefont {G.~J.}\
  \bibnamefont {MacDougall}}, \bibinfo {author} {\bibfnamefont {T.~C.}\
  \bibnamefont {Chiang}}, \bibinfo {author} {\bibfnamefont {E.}~\bibnamefont
  {Fradkin}},  \emph {et~al.},\ }\bibfield  {title} {\bibinfo {title}
  {Signatures of exciton condensation in a transition metal dichalcogenide},\
  }\href@noop {} {\bibfield  {journal} {\bibinfo  {journal} {Science}\ }\textbf
  {\bibinfo {volume} {358}},\ \bibinfo {pages} {1314} (\bibinfo {year}
  {2017})}\BibitemShut {NoStop}%
\bibitem [{\citenamefont {Seki}\ \emph {et~al.}(2014)\citenamefont {Seki},
  \citenamefont {Wakisaka}, \citenamefont {Kaneko}, \citenamefont {Toriyama},
  \citenamefont {Konishi}, \citenamefont {Sudayama}, \citenamefont {Saini},
  \citenamefont {Arita}, \citenamefont {Namatame}, \citenamefont {Taniguchi}
  \emph {et~al.}}]{seki2014excitonic}%
  \BibitemOpen
  \bibfield  {author} {\bibinfo {author} {\bibfnamefont {K.}~\bibnamefont
  {Seki}}, \bibinfo {author} {\bibfnamefont {Y.}~\bibnamefont {Wakisaka}},
  \bibinfo {author} {\bibfnamefont {T.}~\bibnamefont {Kaneko}}, \bibinfo
  {author} {\bibfnamefont {T.}~\bibnamefont {Toriyama}}, \bibinfo {author}
  {\bibfnamefont {T.}~\bibnamefont {Konishi}}, \bibinfo {author} {\bibfnamefont
  {T.}~\bibnamefont {Sudayama}}, \bibinfo {author} {\bibfnamefont
  {N.}~\bibnamefont {Saini}}, \bibinfo {author} {\bibfnamefont
  {M.}~\bibnamefont {Arita}}, \bibinfo {author} {\bibfnamefont
  {H.}~\bibnamefont {Namatame}}, \bibinfo {author} {\bibfnamefont
  {M.}~\bibnamefont {Taniguchi}},  \emph {et~al.},\ }\bibfield  {title}
  {\bibinfo {title} {Excitonic {Bose-Einstein} condensation in {Ta$_2$NiSe$_5$}
  above room temperature},\ }\href@noop {} {\bibfield  {journal} {\bibinfo
  {journal} {Physical Review B}\ }\textbf {\bibinfo {volume} {90}},\ \bibinfo
  {pages} {155116} (\bibinfo {year} {2014})}\BibitemShut {NoStop}%
\bibitem [{\citenamefont {Lu}\ \emph {et~al.}(2017)\citenamefont {Lu},
  \citenamefont {Kono}, \citenamefont {Larkin}, \citenamefont {Rost},
  \citenamefont {Takayama}, \citenamefont {Boris}, \citenamefont {Keimer},\
  and\ \citenamefont {Takagi}}]{lu2017zero}%
  \BibitemOpen
  \bibfield  {author} {\bibinfo {author} {\bibfnamefont {Y.}~\bibnamefont
  {Lu}}, \bibinfo {author} {\bibfnamefont {H.}~\bibnamefont {Kono}}, \bibinfo
  {author} {\bibfnamefont {T.}~\bibnamefont {Larkin}}, \bibinfo {author}
  {\bibfnamefont {A.}~\bibnamefont {Rost}}, \bibinfo {author} {\bibfnamefont
  {T.}~\bibnamefont {Takayama}}, \bibinfo {author} {\bibfnamefont
  {A.}~\bibnamefont {Boris}}, \bibinfo {author} {\bibfnamefont
  {B.}~\bibnamefont {Keimer}}, and\ \bibinfo {author} {\bibfnamefont
  {H.}~\bibnamefont {Takagi}},\ }\bibfield  {title} {\bibinfo {title} {Zero-gap
  semiconductor to excitonic insulator transition in {Ta$_2$NiSe$_5$}},\
  }\href@noop {} {\bibfield  {journal} {\bibinfo  {journal} {Nature
  communications}\ }\textbf {\bibinfo {volume} {8}},\ \bibinfo {pages} {1}
  (\bibinfo {year} {2017})}\BibitemShut {NoStop}%
\bibitem [{Note4()}]{Note4}%
  \BibitemOpen
  \bibinfo {note} {The use of the term condensate is usually taken to mean that
  there is an analogy to Bose-Einstein condensation, but it should be kept in
  mind that this analogy is always imprecise since the number of electrons and
  holes are not conserved separately in bulk three dimensional
  crystals.}\BibitemShut {Stop}%
\end{thebibliography}%

\end{document}